\documentclass{article}

\def\giorno{23/2/2004}

\def\*{{\bf ***}}

\def\a{\alpha}
\def\b{\beta}

\def\ga{\gamma}
\def\de{\delta}   
\def\eps{\varepsilon}
\def\phi{\varphi}

\def\s{\sigma}

\def\om{\omega}
\def\th{\theta}
\def\vth{\vartheta}

\def\C{{\bf C}}
\def\D{{\cal D}}

\def\G{{\cal G}}

\def\J{{\cal J}}

\def\L{{\cal L}}

\def\N{{\cal N}}

\def\R{{\bf R}}

\def\T{{\rm T}}
\def\V{{\cal V}}
\def\W{{\cal W}}

\def\Ga{\Gamma}
\def\De{\Delta}
\def\La{\Lambda}
\def\Om{\Omega}

\def\pa{\partial}

\def\d{{\rm d}}       
\def\w{\wedge}
\def\xb{{\bf x}}

\def\o+{\oplus}

\def\ss{\subset}
\def\sse{\subseteq}

\def\<{\langle}
\def\>{\rangle}

\def\interno{\hskip 2pt \vbox{\hbox{\vbox to .18
truecm{\vfill\hbox to .25 truecm
{\hfill\hfill}\vfill}\vrule}\hrule}\hskip 2 pt}

\def\({\left(}
\def\){\right)}
\def\[{\left[}
\def\]{\right]}
\def\=#1{\bar #1}
\def\~#1{\widetilde #1}
\def\.#1{\dot #1}
\def\^#1{\widehat #1}
\def\"#1{\ddot #1}

\def\ref#1{\cite{#1}}

\def\EOP{~\hfill $\triangle$}
\def\EOD{~\hfill $\oslash$}
\def\EOS{~\hfill $\diamondsuit$}
\def\EOR{~\hfill $\odot$}

\begin{document}

\title{\bf Variational principles for involutive systems of vector fields}

\author{Giuseppe Gaeta\footnote{e-mail: giuseppe.gaeta@mat.unimi.it ; g.gaeta@tiscali.it} \\
{\it Dipartimento di Matematica, Universit\`a di Milano} \\
{\it via Saldini 50, I--20133 Milano (Italy)} \\ {~~} \\
Paola Morando\footnote{e-mail: paola.morando@polito.it} \\
{\it Dipartimento di Matematica, Politecnico di Torino} \\
{\it Corso Duca degli Abruzzi 24, I--10129 Torino (Italy)} }

\date{\giorno }

\maketitle

\noindent {\bf Summary.} In many relevant cases -- e.g., in
hamiltonian dynamics -- a given vector field can be characterized
by means of a variational principle based on a one-form. We
discuss how a vector field on a manifold can also be characterized
in a similar way by means of an higher order variational
principle, and how this extends to involutive systems of vector
fields.
\bigskip

\section*{Introduction}

The paradigm of {\it a vector field identified by a variational
principle} comes from Mechanics, and takes the form of the
Euler-Lagrange or Hamilton equations, depending on the formulation
of the theory.

In these cases, a vector field on a manifold $P$ (in the
Lagrangian formulation, $P = TV$ with $V$ the configuration space;
in the Hamiltonian one, $P = T^* V$), termed the {\it phase
space}, is identified in terms of a variational principle defined
by a one-form on a fiber bundle having the {\it extended phase space}
$M = P \times \R$ as total space, with base the factor $\R$
corresponding to physical time.

More generally, consider a $n$-dimensional bundle $(M,\pi,B)$ on a $k$-dimensional manifold $B$; denote by $\Ga (\pi)$ the set of smooth sections of this bundle. If $\vth$ is a $k$-form on $M$ satisfying certain non-degeneration conditions (depending on the fibration $\pi$), and $D$ any given domain in $B$, we consider for any $\phi \in \Ga (\pi)$ the integral
$$ I [ \phi ] \ := \ \int_D \phi^* ( \vth ) \ ; $$
thus $I$ identifies a smooth real function $I : \Ga (\pi) \to \R$. The request that for a variation of $\phi$ of order $\eps$, vanishing at $\pa D$, the variation of $I$ is of order $o (\eps )$, defines a variational problem (see below for more precise statements, concerning this and other concepts mentioned in this introduction).

Consider first the case where $B = \R$. If the equations expressing the condition $\de I / \de \phi = 0$ identifies a section $\phi$ which is the integral line of a vector field $X$, we say that $X$ is identified (up to normalization) by the variational principle given by $\vth$.

For $B$ higher dimensional, we would obtain field equation, and
the critical sections would be $k$-dimensional submanifolds of $M$.
It was shown in \cite{GMvar} that for $k = n-2$ -- i.e. for the
higher possible degree of $\vth$, see below -- these manifolds are
actually integral manifolds of a one-dimensional module of vector
fields; that is, in this case as well the variational principle
identifies (up to normalization) a vector field $X$.

\medskip\noindent
{\bf Remark 1.} It is important to stress that a variational
principle by itself will always identify a {\it module} (over
$\C^\infty (M)$) of vector fields rather than a single one; in
order to single out a specific vector field from this module, one
needs an additional requirement; usually this is simply a
normalization condition. \EOR
\medskip

The purpose of this note is twofold: on the one hand we want to
illustrate how a vector field can also be identified by a maximal
order ($k = n-2$ in our present notation) variational principle,
as proven in recent work \cite{GMlio,GMvar}; on the other hand we
want to discuss if, under suitable conditions,  {\it a variational
characterization is also possible for systems of vector fields in
involution} (rather than a single vector field); we will answer
this question in the positive and identify this with $\N (\d \vth )$, i.e. with the characteristic distribution of the variational ideal $\J (\vth , \pi )$, see below. We will not try to give a general
discussion, but just study a special class of forms $\vth$: those for which $\d \vth$ is a decomposable form satisfying certain nondegeneracy conditions.

The point raised in remark 1 will also be relevant here: that is,
the variational principle by itself will identify a module of
vector fields rather than a finite dimensional set; we can reduce
to the latter only by additional conditions.

It turns out that the convenient language to discuss this problem
is provided by the theory of {\it Cartan ideals}; we will actually
to a large extent make use of the framework laid down in
\cite{GMvar}, adapting it to our present purposes.

Sections 1-3 will be devoted to illustrate this framework as well
as (the parts we need of) classical Cartan ideals theory. In
section 4 we discuss the relation between the theory developed in
previous section and {\it reduction} (in the sense of proposition
4, see sect.1), i.e. how a variational principle based on a
$k$-form (with $k>1$) on a $n$-dimensional manifold, which of
course produces a system of PDEs and provides critical sections
$\s$ corresponding to $k$-dimensional submanifolds of $M$, can
also identify a ($q = n-k-1$ dimensional) module of vector fields.
The determination of critical sections $\s$ can then be reduced to
determining a $(k-q)$-dimensional manifold $\s_0$ which is in a
way the quotient of $\s$ by the action of the vector fields.
Section 5 will recall the results that are obtained in the {\it
``maximal degree''} case $k = n-2$, where the module is
one-dimensional -- i.e. imposing a normalization condition we have
a single vector field --   while section 6 will deal with the {\it
decomposable} case (we defer consideration of more general cases
to a separate work). In section 7 we briefly discuss, for the sake
of completeness, the case of non-proper variational principles. In
the last three sections we provide completely explicit examples,
dealing respectively with the ``maximally characteristic'' and the
``non maximally characteristic'' cases (see sect.6), and with a
``non proper'' variational principle.

\medskip\noindent
{\bf Acknowledgements.} The work of GG was supported in part by
{\it ``Fondazione CA\-RI\-PLO per la ricerca scientifica''}. We
thank prof. G. Sardanashvily for his invitation to write this
paper for the inaugural volume of IJGMMP.

\section{Cartan ideals}

In this section we will recall some basic notions from the theory
of {\it Cartan ideals}, i.e. ideals of differential forms. The
reader is referred to \cite{Car} for further detail, and
\cite{Arn,Bry,Olv} for modern expositions and further
developements.\footnote{See e.g.  \cite{Arn,God} for the use of
Cartan's ideals in the study of PDEs and in analytical mechanics,
including standard variational formulation of the latter. The
relation between Cartan ideals and variational problem is studied
in great detail, for $B$ one dimensional, in \cite{Gri}. The
geometry of PDEs is naturally discussed using Cartan ideals, see
e.g. \cite{AVL,Arn,Bry,Car,KrV,Olv}.}

From now on $M$ will be a smooth $n$-dimensional manifold; we will denote by $i$ the canonical inclusion, so that a submnaifold $S \ss M$ will also be denoted by $i : S \to M$.

\medskip\noindent
{\bf Definition 1.} We say that $\J \ss \Lambda (M)$ is a {\bf Cartan ideal} iff: {\bf (i)} it is an ideal in $\Lambda (M)$ under exterior product; {\bf (ii)} $\J_k := \J \cap \Lambda^k (M)$ is a module over $\Lambda^0 (M)$ for all $k =0,...,n$. These are also rephrased as follows: {\bf (i)} for all $\eta \in \J$, $\psi \in \Lambda (M)$, $\eta \w \psi \in \J$; and {\bf (ii)} for  all $\beta_i \in \J_k$, $f_i \in \Lambda^0 (M)$ ($i=1,2$), $f_1 \beta_1 + f_2 \beta_2 \in \J_k$ (for all $k=0,...,n$). \EOD

\medskip\noindent
{\bf Definition 2.} Let $i : S \to M$ be a smooth submanifold of $M$; $S$ is said to be an {\bf integral manifold} of the Cartan ideal $\J$ iff $i^* (\eta ) = 0$ for all $\eta \in \J$. In other words, $S \ss M$ is an integral manifold of $\J$ iff all $\eta \in \J$ vanish on $S$. \EOD
\medskip

The Cartan ideal $\J$ is said to be {\it generated} by the forms $\{ \eta^{(\a)} , \a= 1,...,r \}$ (with $\eta^{(\a)} \in \J$) if each $\zeta \in \J$ can be written as $\zeta = \sum_\a \rho_{(\a)} \w \eta^{(\a)}$ for a suitable choice of $\rho_{(\a)} \in \Lambda (M)$, $\a = 1,...,r$.

\medskip\noindent
{\bf Proposition 1.} {\it If $\J$ is generated by $\{ \eta^{(\a)} , \a= 1,...,r \}$, then $i:S \to M$ is an integral manifold for $\J$ iff $i^* (\eta^{(\a)}) = 0$ for all $\a = 1,...,r$.} \EOS
\medskip

The Cartan ideal $\J$ is said to be {\it closed} if it is closed under exterior differentiation, i.e. if $\d \eta \in \J$ for all $\eta \in \J$. In this case one also says that $\J$ is a {\it differential ideal}.

If the Cartan ideal $\J$ is generated by $\{ \eta^{(\a)} , \a= 1,...,r \}$, it can always be completed to a differential ideal by adding the  $\d \eta^{(\a)} \not\in \J$ to the system of generators. We denote by $\widehat{\J} $ the completion of the ideal $\J$ obtained in this way; obviously $\J \sse \widehat{\J}$, the equality corresponding to the case where $\J$ is closed.

Note that if $\eta$ vanishes on $S$, the same is true of $\d \eta$;
thus, integral manifolds of $\J$ are also integral manifolds of
$\widehat{\J}$.\footnote{In Cartan's words, ``La recherche des
solutions d'un syst\`eme diff\'erentiel peut toujours \'etre
ramen\'ee \`a la recherche des solutions d'un syst\`eme
diff\'erentiel ferm\'e '' (see \cite{Car}, p. 52).} We will always
assume that $\J$ does not include 0-forms; by the previous remark,
this is not actually a limitation (but simplifies discussions).

Note that if $\eta = \d \a$ and $i : S \to M$, then $i^* (\eta ) = 0$
means that $\a$ is constant on $S$. In particular, if we deal with a
set of equations $F_a = 0$ on $M$, we can pass to the  system $\eta_a
:= \d F_a = 0$; integral manifolds for the (closed) Cartan ideal
generated by the $\eta_a$ will be manifolds on which the $F_a$ are
constant; the solution to the original problem will be provided by
the manifold on which they are constant and all equal to zero.
\bigskip

Given a Cartan ideal $\J$, we associate to any point $x \in M$ the subspace $D_x (\J) \ss \T_x M$ defined by
$$ D_x (\J) \ := \ \{ \xi \in \T_x M  : \, \xi \interno \J_x \ss \J_x \} \ . $$
If $D_x (\J)$ has constant dimension, the Cartan ideal $\J$ is said to be {\it non singular}, and the distribution $D (\J) = \{ D_x (\J) , x \in M\}$ is its {\it characteristic distribution}; any vector field $X \in D (\J)$ (by this we mean that $X (x) \in D_x (\J )$ at all points $x \in M$) is said to be a {\it characteristic field} for $\J$.

\medskip\noindent
{\bf Remark 2.} Note that if all the generators $\eta^{(\a)}$ of $\J$ are of the same degree $k$,  then all forms in $\J$ are of degree not smaller than $k$, and $\J_m = \{ 0 \}$ for $m < k$. If $\J_m = \{ 0 \} $ for $m<k$, then $X \in D (\J)$ satisfies $X \interno \zeta = 0$ for all $\zeta \in \J_k$, and in particular $X \in D (\J)$ iff $X \interno \eta^{(\a)} = 0$. Indeed by definition any $\zeta \in \J$ is written as $\zeta = \rho_{(\a)} \w \eta^{(\a)}$, and $X \interno \zeta = \s_{(\a)} \w \eta^{(\a)}$ with $\s_{(\a)} = X \interno \rho_{(\a)}$. \EOR

\medskip\noindent
{\bf Definition 3.} An {\bf integral manifold} for a distribution $D$
on $M$ is a submanifold $i:N \to M$ such that $i_* (\T_x N ) \ss D_{i
(x)}$ for all $x \in N$. In other words, any vector field tangent to
$N$ is in  $D$ (the converse is in general not true). \EOD
\medskip

It should be stressed that integral manifolds of $D(\J)$ are always
integral manifolds of $\J$, but the converse is in general not true.

\medskip\noindent
{\bf Definition 4.}
The $p$-dimensional distribution $D$ on $M$ is said to be {\bf completely integrable} if through each point $x \in M$ passes a $p$-dimensional integral manifold of $D$. In this case, the $p$-dimensional integral submanifolds are also said to be the {\it Cauchy characteristics} for $D$. \EOD

\medskip\noindent
{\bf Proposition 2.} {\it If $\J$ is a closed nonsingular differential Cartan ideal, then $D (\J)$ is completely integrable.} \EOS
\medskip

It should be stressed that the Cauchy characteristics of an integrable  $p$-dimensional distribution $D$ provide a foliation of $M$ by $p$-dimensional submanifolds \cite{Nar}. Thus if $\J$ is a closed nonsingular Cartan ideal with $p$-dimensional characteristic distribution $D (\J)$, then $\J$ always has $p$-dimensional integral manifolds, and $M$ is foliated by these (see below the notion of
complete ideal).

The following theorem (proposition 3) is most useful in performing computations with Cartan ideals; it appears in different forms in \cite{Bry,Car,Olv}. Proposition 4 is an immediate consequence of it; see section 45 of \cite{Car}.

\medskip\noindent
{\bf Proposition 3.} {\it Let $\J$ be a nonsingular differential
Cartan ideal, and let its characteristic distribution $\D (\J)$ be
$p$-dimensional. Then in a neighbourhood of any point $x \in M$ we
can choose local coordinates $(x^1 , ... , x^p ; y^1 , ... ,
y^{n-p})$ such that $\J$ admits a system of generators $\{ \th_1 ,
... , \th_r \}$ with the property that, locally around $x$, the
$\th$ and $\d \th$ do not involve the variables $x^j$ nor the forms $\d x^j$.} \EOS
\medskip

The local coordinates whose existence is guaranteed by this theorem
will be called {\it Cartan canonical coordinates}; if we consider
locally a fibration of $M$ over $\R^p$ for which the $x^i$ are
horizontal and the $y^j$ are vertical coordinates, $\D (\J)$ spans
horizontal planes identified as $y^j = const$, $j=1,...,n-p$.

\medskip\noindent
{\bf Proposition 4.} {\it  Let $\J$ be a nonsingular differential
Cartan ideal, and let $\G$ be the $p$-dimensional characteristic
distribution for $\J$; let $i: S \to M$ be a $q$-dimensional
integral manifold of $\J$. Assume that $\G$ is nowhere tangent to
$i (S)$, and denote by $G(x)$ the local integral manifold for $\G$
through a point $x$. The $(p+q)$-dimensional local manifold $\Phi : G(S) \to M$ defined by the union of the $G(x)$ through points in $S$ is a local integral manifold of $\J$.} \EOS
\medskip

Finally, let us consider the useful notion of the {\it
complete ideal} (sometimes also called characteristic ideal), see \cite{Sch}, related to a  Cartan ideal.
Consider the ideal $\J$ and its characteristic distribution $D
(\J)$. The complete ideal $\bar{\J}$ is the set of forms
$\om \in \La (M)$ which are annihilated by all vectors in $D(\J)$,
i.e. $$ \bar{\J} \ := \ \{ \om \ : \ X \interno \om = 0 \ \forall
X \in D (\J) \} \ ; $$ note that this can and in general (i.e.
unless $\J$ is generated by a set of one-forms) will include forms
of degree lower than those in $\J$.

The integrability of $\D (\J)$ can be studied by means of the forms
$\a_i$ generating $\bar{\J}$ (this is just another version of
Frobenius theorem, see \cite{Sch}).

The complete ideal $\bar{\J}$ can always be generated -- as
a  Cartan ideal -- by a set of one-forms $\a_i \in \La^1 (M)$. In
the case of interest here, i.e. for a non-singular ideal $\J$,
these are easily built as follows: if $\{ X_1 , ... , X_p \} $ are
vector fields spanning $D (\J) $ as a module, complete the set by
any set of vectors $\{Y_1 , ... , Y_{n-p} \}$ such that the $\{
X_i ; Y_j \} $ together span $T M$, and choose these so that $(X_i
\cdot Y_j ) = 0$ for all $i,j$. Then the $\a_i$ are the one-forms
dual to the $Y_i$.

\section{Variational principles and variational modules}

In this section we recall the construction of variational modules given in \cite{GMvar} (see there for further detail), and its relation to standard notions in the calculus of variations.

Let $\pi : M \to B$ be a smooth bundle; we assume that $M$ is $n$-dimensional, and $B$ is a smooth manifold of dimension $k$, with $1 \le k < n$.

We denote, as customary, by $\Ga (\pi )$ the set of smooth sections of the bundle $\pi : M \to B$, and by $\V (\pi)$ the set of vector fields in $M$ which are vertical for this fibration. For $D$ a domain in $B$, we denote by $\V_D (\pi) \ss \V (\pi)$ the set of vertical vector fields which vanish on all of $\pi^{-1} (\pa D)$. We will use such notations for all bundles.

Consider a form $\vth \in \Lambda^k (M)$ (not basic for the
fibration $\pi$); then to any domain $D \ss B$ we associate a
functional $I_D : \Ga (\pi ) \to \R$ by $$ I_D (\phi) \ := \
\int_D \phi^* (\vth ) \ . \eqno(1) $$ Let $V\in \cal V(\pi)$ and
$\gamma \in \Gamma (\pi)$; denote by $\psi_s$ the flow of $V$ on
$M$. This induces a flow in $\Gamma$, and the flow of $\gamma$ is
the one-parameter family of local sections $\widetilde{\psi}_s
(\gamma):= \psi_s \circ\gamma$. The variation under $V$ of $I_D$
at $\phi \in \Ga (\pi )$ is defined as $$ (\de_V I_D ) (\phi) \ :=
\ {\d ~ \over \d s} \ \[ \int_D \, \( \widetilde{\psi}_s (\phi)
\)^* (\vth) \]_{s=0} \ . \eqno(2) $$

The requirement that $(\de_V I_D ) (\phi ) = 0$ for all $V \in \V_D (\pi )$ [we write $\de I_D (\phi)$ for short] is the {\it variational principle on $\pi : M \to B$ defined by $\vth$}. With reference to the degree of $\vth$ (equal to the dimension of $B$), we say this is a {\it variational principle of degree $k$}. If $\d \a = 0 $, then the variational principle defined by $\vth' = \vth + \a$ is equivalent to the one defined by $\vth$.

We want to exclude the possibility that the variation of $\vth$ be identically zero along some vertical direction; this leads us to introduce the notion of proper variational principle.

\medskip\noindent
{\bf Definition 5.} The variational principle on $\pi:M \to B$
defined by $\vth$ is {\bf proper} if $\d \vth$ is nowhere zero and
there is no vertical field along which the variation is zero for
all sections, i.e. there is no vertical field $V \in \V (\pi)$
such that $V \interno \d \vth = 0$. ~~~~\EOD
\medskip

A section $\phi \in \Gamma (\pi)$ is {\it critical for $I_D$} if
and only if $(\de_V I_D ) (\phi) = 0 $ whenever $V \in {\cal V}_D
(\pi)$. A well known criterion for a section to be critical is as
follows (see e.g. \cite{Her}).

\medskip\noindent
{\bf Proposition 5.} {\it A section $\phi \in \Gamma (\pi)$ is
critical for $I_D$ if and only if $ \phi^* ( V \interno \d \vth )
=  0$ for all  $V \in \V_D (\pi)$.} \EOS
\medskip

We introduce now the concept of variational module, and provide an equivalent criterion for $\phi$ to be critical in terms of this \cite{GMvar}.

Consider a basis $\{ V_1 , ... , V_r \} $ (here and below, $r=n-k$) of vertical vector fields, generating $\V (\pi )$ as a module. Then any $V \in \V (\pi )$ can be written as $V = \sum_{i=1}^r f^i (x) V_i $, and $V \in \V_D (\pi ) \ss \V (\pi)$ if and only if $f^i (x) = 0$ for all $x \in \pi^{-1} (\pa D)$ and for all $i=1,...,r$.

Define the forms $\Psi_j \in \Lambda^k (M)$ as $\Psi_j := V_j
\interno \d \vth$ (for $j=1,...,r$). The module $\W (\pi,\vth )$
generated by $\{ \Psi_1 , ... , \Psi_r \}$ is the {\it variational
module} associated to the variational principle over $\pi : M \to B$
defined by $\vth$. Note that $\W (\pi , \vth)$ does not depend on the
choice of the basis $\{ V_j \}$; moreover, the variational modules
for $\vth$ and for $\vth' = \vth + \b$ with $\b$ closed, are
equivalent. If $\W (\vth,\pi)$ is $r$-dimensional, we say it is {\it
nondegenerate}.

We can then rephrase proposition 5 as follows \cite{GMvar} (note
that this condition is manifestly independent of $D$):

\medskip\noindent
{\bf Proposition 6.} {\it A section $\phi \in \Gamma (\pi)$ is
critical for $I_D$ if and only if $\phi^* (\W) =0$, i.e. iff
$\phi^* (\Psi) = 0 $ for all $\Psi \in \W (\pi,\vth)$.} \EOS
\medskip

In studying the variational principle defined by $\vth$, a central
role is played by the annihilator of $\d \vth$. Let us hence
consider the annihilator $\N (\eta)$ of a form $\eta \in
\Lambda^{k+1} (M)$, i.e. the module of vector fields $Y$ on $M$
such that $Y \interno \eta = 0$. It is shown in \cite{GMvar} that
if $\{X;V_1,...,V_r \}$ are $n-k+1$ independent and nonzero vector
fields on $M$ ($r=n-k$), then $V_j \interno (X \interno \eta) = 0$
for all $j=1,...,r$ is equivalent to $X \interno \eta = 0$.

By specializing to $\eta = \d \vth$, this implies that if $\vth
\in \Lambda^k (M)$ is non closed (and non basic for $\pi : M \to
B$), then a vector field $X \not\in \V (\pi) $ satisfies $X
\interno \W (\vth,\pi) = 0$ iff $X \in \N (\d \vth)$.

In other words, the set of vector fields which are transversal to
the fibers of $\pi$ and annihilate $\W (\vth,\pi)$ corresponds to
the set of vector fields in $\N (\d \vth)$ which are not vertical.

Note that if $\vth$ defines a proper variational principle in
$\pi:M \to B$, then (by definition 5) $\N (\d \vth )$ will not
contain any vector field which is vertical for $\pi$. We thus have
the

\medskip\noindent
{\bf Lemma 1.} {\it Let $\vth$ define a proper variational
principle in $(M,\pi,B)$. Then the characteristic distribution $\D
= D [\J (\vth,\pi)]$ of the Cartan variational ideal $\J
(\vth,\pi)$ coincides with the distribution $\N (\d \vth)$.} \EOS

\medskip\noindent
{\bf Remark 3.} For a generic nonzero $\eta$, we are not
guaranteed that $\N (\eta ) \not= \{ 0 \}$, nor that there are
nonzero independent vectors $\{X;V_1,...,V_r \}$ as above.
Moreover, the rank of $\N_x (\eta ) := \{ \xi \in \T_x M : \eta_x
(\xi)=0 \}$ could be different at different points $x \in M$. \EOR

\section{Cartan ideals and variational principles}

We will now consider the Cartan ideal $\J$ generated by $\W
(\pi,\vth)$; by this we mean the ideal generated by a set of
generators of $\W (\pi,\vth)$, which corresponds to a set of
generators $V_j$ for $\V (\pi)$. Note that, as remarked above,
this does not depend on the choice of the $V_j$, and is invariant
under adding to $\vth$ a closed form.

\medskip\noindent
{\bf Definition 6.} The Cartan ideal  $\J (\vth,\pi)$ generated by
$\W (\vth,\pi)$ is the ``Cartan ideal associated to the
variational principle on $\pi$ defined by $\vth$''. We will refer
to it, for short, as the {\bf variational ideal}. \EOD
\medskip

Note that if $(\d \vth)_{x_0} = 0$ at some point $x_0 \in M$, then
$\Psi_j = \pa_j \interno \d \vth$ also vanish at that point, and
$D_{x_0} (\J) = \T_{x_0} M$. Thus in order to have a nonsingular
$\J (\vth , \pi)$, we have to require that $\d \vth$ is nowhere
zero (if the variational principle is proper, this is
automatically true).

We can characterize critical sections of the variational principle on $\pi : M \to B$ defined by $\vth$ by noting that the critical sections of the variational principle on $\pi : M \to B$ based on $\vth$ are integral manifolds of the Cartan ideal $\J (\vth,\pi)$.
We can therefore rephrase proposition 6 (which was a restatement of proposition 5) in terms of Cartan ideals.

\medskip\noindent
{\bf Proposition 7.} {\it A section $\phi \in \Gamma (\pi)$ is critical for the proper variational principle on $\pi : M \to B$ defined by $\vth$ if and only if $\phi$ is an integral manifold of the Cartan variational ideal $\J (\vth,\pi)$.} \EOS
\medskip

This proposition justifies calling $\J (\vth,\pi)$ the Cartan ideal associated to the variational principle $\de I_D  = 0$: indeed, it implies that in order to study (critical sections for) the variational principle $(\de I_D)(\phi) = 0$, we can just study (integral manifolds of) the Cartan ideal $\J(\vth,\pi )$.

We stress that, more precisely, we have to study integral manifolds of $\J (\vth,\pi)$ that are sections of $\pi : M \to B$; this means in particular that they are of dimension $k$ and everywhere transversal to fibers of the bundle $\pi : M \to B$.

We have thus completely characterized critical sections $\phi$ for a variational principle as sections which are integral manifolds for the associated Cartan variational ideal.

In the previous section, we considered the (necessarily non-vertical, if the variational principle is proper) vector fields $X \in \N (\d \vth )$. These are, by construction, characteristic for the variational ideal $\J (\pi , \vth )$ and will therefore be tangent to its integral manifold, i.e. -- see proposition 7 -- to the critical section for the variational principle.

\section{Variational principles and reduction}

Consider the variational principle on $\pi : M \to B$ defined by
$\vth \in \Lambda^k (M)$; assume $\J := \J (\vth,\pi)$ is
nonsingular, and $\D := D[\J (\vth,\pi)]$ is an integrable
$q$-dimensional distribution.

\medskip\noindent
{\bf Remark 4.} We stress that both these assumptions are non
generic; needless to say, the discussion of this section will apply only under these hypotheses. \EOR
\medskip

The result of proposition 4 can be applied to reduce the problem
of determining critical section of a variational principle, i.e.
$k$-dimensional integral manifolds of $\J (\vth,\pi)$ transversal
to the fibers of $\pi : M \to B$, down to that of determining
$(k-q)$-dimensional ones satisfying suitable transversality
conditions with respect to the fibration $\pi : M \to B$ and also
to the foliation provided by $\D$ (these conditions are
automatically satisfied if $\vth$ defines a proper variational
principle in $\pi : M \to B$).

We say that the submanifold $M_0 \ss M$ is {\it non
characteristic} for $\J$ if $T_x M_0 \cap [D (\J)]_x = \{ 0 \}$
for all $x \in M_0$, i.e. if it is everywhere transversal to the
characteristic distribution of the ideal. Then a local integral
manifold for $\J$ is specified by assigning a manifold $M_0$ which
is integral and non characteristic for $\J$, and ``pulling'' it
along the characteristic distribution $\D$.

In a less pictorial way, we build -- as described in proposition 4
-- a local integral manifold for $\J$ as a local bundle over
$M_0$, with fibers corresponding to integral manifolds for $\D$
(see proposition 2 and the remark after it); note this only uses
the Frobenius integrability of $\D$ \cite{Arn,Car}.

It should be stressed that, of course, such a general reduction is {\it not} always possible; actually when the fibers of $\pi : M \to B$ have dimension greater than two it is generally impossible to perform it (the case of two-dimensional fibers presents several peculiarities also in this respect, see section 5 below), as we now briefly discuss.

When looking for integral manifolds of $\J$ which are sections of
$\pi : M \to B$, this reduction would require to consider the
subset $\D_\pi \sse \D$ which is transversal to fibers of $\pi : M
\to B$, and extend integral manifolds of $\J$ over a submanifold
$B_0 \ss B$ of codimension equal to the dimension of $\D_\pi$ to a
local critical section.

Note that several additional conditions are required for the
reduction procedure to be viable: the dimension of $\D_\pi$ can
vary even if that of $\D$ is constant; moreover, the involutivity
of $\D$ does not imply, in general, involutivity and hence
integrability of $\D_\pi$. In practice, this means that this
approach can be applied to the construction of critical sections, i.e. integral manifolds of $\J$ which are sections of the bundle $\pi : M \to B$, only if $D [\J (\vth,\pi)]$ is transversal to the fibers of $\pi : M \to B$; that is, there are nondegeneracy conditions which must be satisfied by $\vth$ or equivalently by $\W (\vth,\pi)$. These are automatically satisfied when $\J$ is the variational ideal for a proper variational principle in $\pi : M \to B$.

An even more substantial obstacle is that $\N (\d \vth )$ (and thus the ``useful'' part of $D (\J)$, see section 2) is in general empty when $\vth$ does not have degree $k = n-2$ (see \cite{GMvar} for a discussion of the special features of the latter case).
In this case, of course, we miss the main ingredient of the reduction procedure.

\medskip\noindent
{\bf Remark 5.} The above discussion can be better reinterpreted
in terms of the Cartan canonical coordinates (see proposition 3).
We work in $\pi : M \to B$ and look for integral manifolds of a
Cartan ideal $\J$ which are sections for $\pi$. The Cartan
coordinates define a (local) natural fibration $\kappa : M \to L$
over a $p$-dimensional manifold $L$, spanned (in the notation of
proposition 3) by the coordinates $x^1 , ... , x^p$.

Thus we have two local fibrations in $M$, i.e. $\pi : M \to B$ and
$\kappa : M \to L$. The latter is such that $D (\J)$ is
transversal to fibers $\kappa^{-1} (\ell)$ for all $\ell \in L$,
but in order to apply the reduction procedure to integral
manifolds of $\J$ which are sections of $\pi : M \to B$, we need
that $D(\J)$ be transversal to fibers $\pi^{-1} (b) $ for all $b
\in B$.
This condition is in general {\it not}  satisfied, but it is
automatically met when $\J = \J (\vth , \pi)$ is the variational
ideal for a proper variational principle in $(M,\pi,B)$. \EOR
\medskip

\section{The maximal degree case}

In the maximal degree case\footnote{If $k = n-1$, we have $\d \vth
\in \Lambda^n (M)$, so it is either degenerate or a volume form;
in both cases it does not define a proper variational principle.},
i.e. for $\vth \in \La^k (M)$ with $k = n-2$, the nondegenerate
form $\eta := \d \vth$ is of degree $n-1$; it is well known that
in this case $\N (\d \vth )$ is necessarily a one-dimensional
module. This implies that our general construction applies here,
as we discuss in this section.

\subsection{Abstract results}

In this case our general discussion concretizes into the following results, see \cite{GMvar}.

\medskip\noindent
{\bf Proposition 8.} {\it Let $\pi : M \to B$ be a smooth fiber
bundle of dimension $n$ with base manifold $B$ of dimension $k= n-2$;
let $\vth \in \Lambda^k (M)$ be non basic for this fibration, and
such that $\eta := \d \vth$ is nowhere zero on $M$. Then the Cartan
ideal $\J (\vth,\pi)$ is nonsingular and admits a one-dimensional
characteristic distribution $D[\J (\vth,\pi )]$; this coincides with
$\N (\d \vth )$.} \EOS
\medskip

Note that vector fields in $D [\J (\vth,\pi)]$ differ only by a nonzero function, and can thus be uniquely determined by a normalization prescription (see remark 1 above). Thus a maximal degree proper variational principle over $\pi : M \to B$ together with a normalization condition determine a unique vector field in $M$.

In this case one can also apply the reduction procedure discussed above:

\medskip\noindent
{\bf Proposition 9.} {\it Let $B_0 \ss B$ be a smooth submanifold of
codimension one in $B$, and $\pi_0 : \pi^{-1}(B_0) \to B_0$ the
associated subbundle\footnote{This is defined by $M_0 = \pi^{-1}
(B_0) \ss M$, with $\pi_0$ the restriction of $\pi$ to $M_0$;
sections of this subbundle will be denoted as $\Ga (\pi_0 )$.} of
$\pi: M \to B$. Let $\phi_0 \in \Ga (\pi_0)$, seen as a submanifold
of $M$, be an integral manifold for the Cartan ideal $\J (\vth,\pi)$,
nowhere tangent to integral manifolds of $D[\J (\vth,\pi)]$. Then the
critical local sections for the maximal degree variational principle
on $\pi$ defined by $\vth$ can be built by pulling $\phi_0$ along
integral curves of $D [\J (\vth,\pi)]$.} \EOS
\medskip

Finally, let us also consider the inverse problem: given a vector field $X$ on $M$, characterize it, up to normalization, in terms of a maximal degree variational principle (see below for the case of Liouville vector fields).

\medskip\noindent
{\bf Proposition 10.} {\it Let $M$ be a smooth $n$-dimensional manifold, and $X$ a vector field on $M$. Assume there is an exact form $\eta = \d \vth \in \Lambda^{n-1} (M)$ such that: (i) $X \in \N (\eta)$ (ii) $\eta$ is nowhere vanishing, (iii) $\eta$ is not basic for the fibration $\pi : M \to B$ over a $(n-2)$-dimensional manifold $B \ss M$. Then $X$ generates the characteristic distribution of the Cartan ideal associated to the (maximal degree) variational principle on $\pi : M \to B$ defined by $\vth$.} \EOS
\medskip

\subsection{Coordinate approach}

It is worth discussing -- also in view of later extensions -- how the above abstract results are embodied in concrete computations using local coordinates \cite{GMlio}; this will also help to make contact with our general framework. We will work (locally) in euclidean $\R^n$.

We will take coordinates $ \{ x^1 , ... , x^k \}$ on $B$, and $\{ y^1 , ... , y^p \}$ on the fiber. As $k=n-2$, we have $p=2$; we will write $z \equiv y^1$, $w \equiv y^2$ to avoid a plethora of indices.

We write $\om = \d x^1 \w ... \w \d x^k$ for the reference volume
form in $B$; the reference volume form in $M$ will of course be
$\pi^* (\om) \wedge \d z \wedge \d w$; in the following we will write,
with a slight abuse of notation, $\om$ for $\pi^* (\om)$.

One should focus on $\eta := \d \vth \in \Lambda^{n-1} (M)$; we can
always write any $\eta \in \Lambda^{n-1} (M)$ in the form $$
\begin{array}{rl} \eta \ =& \ \sum_{\mu=1}^k A^\mu \[ \om_{(\mu)} \w
\d z \w \d w \] \ + \\ & \ + \ (-1)^k f \[ \om \w \d w \] \ + \
(-1)^{k+1} g \[ \om \w \d z \] \ , \end{array} $$ with
$\mu=1,2$, $A^\mu,f,g$ smooth functions of $({\bf x},z,w)$, and
$\om_{(\mu)} := \pa_\mu \interno \om$.

In the following, we will assume that the vector ${\bf A} = (A^1 ,
... , A^k)$ is not identically zero (if this was the case,
the variational principle would not be proper).

We choose $\pa_z$ and $\pa_w$ as generators of $\V (\pi)$, i.e. $\Psi_1 = \pa_z \interno \eta$, $\Psi_2 = \pa_w \interno \eta$. With $\phi \in \Ga (\pi)$, we have
$$ \begin{array}{rl}
\Psi_1 \ =& \ (-1)^{k-1} \, \[ A^\mu \, (\om_{(\mu)} \w \d w ) \ + \ (-1)^k g \,  \om \] \ ; \\
\Psi_2 \ =& \ (- 1)^k \, \[ A^\mu \, ( \om_{(\mu)} \w \d z) \ + \ (-1)^k f \, \om \] \ . \\
\phi^* (\Psi_1) \ =& \  \phi^* \[ A^\mu (\pa w / \pa x^\mu) \, - \,  g \] \, \om \ ; \\
\phi^* (\Psi_2) \ =& \ - \, \phi^* \[ A^\mu (\pa z / \pa x^\mu) \, - \, f \] \, \om \ . \end{array}  $$

Requiring the vanishing of both $\phi^* (\Psi_j)$ for $j=1,2$ means looking for solutions of two quasilinear first order PDEs; writing
$Y = A^\mu \pa_\mu$, and with $\L_Y$ the Lie derivative, these are
$$ \phi^* \[ \L_Y (z) \, - \, f \] \ = \ 0  \ \ \ ; \ \ \ \phi^* \[ \L_Y (w) \, - \, g \] \ = \ 0 \ . \eqno(3) $$

The relevant property is that the equations can be written in terms
of the action of the same (nonzero) vector field $Y$, or more
precisely \cite{Arn} in terms of the (non vertical, as $Y \not= 0$)
vector field $W = Y + f \pa_z + g \pa_w$ on $M$, i.e. $$ W \ = \
\sum_{\mu=1}^{n-2} \, A^\mu (\xb;z,w) \, {\pa \over \pa x^\mu } \ + \
f (\xb;z,w) \, {\pa \over \pa z} \ + \ g (\xb;z,w) \, {\pa \over \pa
w} \ . \eqno(4) $$

Indeed, see e.g. \cite{Arn}, the $\R^2$-valued function $u(x,t) = \left( z(x,t) , w (x,t) \right)$ is a solution to the system of quasilinear PDEs (3) if and only if its graph is an integral manifold for the associated characteristic system
$$  d x^\mu / d s = A^\mu \ , \ d z / d s =  f \ , \ d w / d s  = g \ . $$

This is just the $W$ given above, and it is thus entirely natural to
call $W$ the {\it  characteristic vector field} for the maximal
degree variational principle on $\pi : M \to B$ defined by $\vth$.

Note that we have a one dimensional module of characteristic vector
fields (all differing by multiplication by a nowhere zero smooth
function); these define a unique direction field on $M$ \cite{Arn}.

Summarizing, with the above discussion we have proved that:

\medskip\noindent
{\bf Proposition 11.} {\it The section $\phi \in \Ga (\pi)$ is critical for the maximal degree proper variational principle defined by $\vth$ if and only if it is an invariant manifold of the characteristic vector field $W$, i.e. is foliated by integral lines of $W$.} \EOS
\medskip

If one of the $A^\mu$, say $A^1$, is nowhere zero, we can divide this out from $W$, and obtain a vector field of the form
$ Z := \pa_1 + X$, with $X \interno \d x^1 = 0$ and hence satisfying the normalization condition $\pa_t \interno Z = 1$.

\subsection{The distribution $\N (\d \vth ) $.}

We will discuss in some detail, due to its relevance in our
general reduction procedure, the geometry of the (one-dimensional)
distribution $\N (\d \vth)$ in this case. This is generated by
$W$, as seen above, so we are actually discussing properties of
$W$.

We know that $W$ is tangent to sections $\phi \in \Ga (\pi)$ such
that $\phi^* (\Psi_1) = 0 = \phi^* (\Psi_2)$, see above and
\cite{Arn}; on the other hand, $\phi^* (\Psi_i) = 0$ means that
$\Psi_i$ vanish on vector fields tangent to $\phi$, hence vanish if
evaluated on $W$. This shows that the characteristic vector field
satisfies $$ W \interno \Psi_i \ \equiv \ W \interno V_i \interno \d
\vth \ = \ 0 \ \ \ \  (i=1,2) \ . $$

With local coordinates $(\xb,z,w)$ as before (so $V_1 = \pa_z$,
$V_2 = \pa_w$), consider a vector field $X$ which is nonzero and
non vertical; hence $V_j \interno (X \interno \d \vth) = 0$ for
$j=1,2$ means that $\chi := X \interno \d \vth$ does not contain
$\d z$ or $\d w$ factors. However, this is impossible unless $X
\interno \d \vth = 0$: {\it The condition $V \interno (X \interno \d \vth) = 0$ for all $V \in \V (\pi)$ implies -- and is thus equivalent to -- $X \interno \d \vth = 0$.}

Indeed, $\chi \in \Lambda^k (M)$, hence we should have $\chi = F(\xb,z,w) \d x^1 \w ... \w \d x^k$; this cannot be obtained by $\chi = X \interno \d \vth$ if $X$ is not vertical, i.e. $X = X_0
+ \b_1 \pa_z + \b_2 \pa_w$ with $X_0 = \a_i \pa / \pa x^i$
nonzero.

In fact, this would mean that either $d \vth= \eta_0 \w
\d z$ or $\d \vth  = \eta_0 \w \d w$, with $\eta_0$ semibasic for
the fibration $\pi: M \to B$; but in this case the variational
principle would not be proper.

Note that this applies to $W$ provided this is non vertical, i.e.
provided the $A^\mu$ identifying $\d \vth$ -- see above -- are not
all identically vanishing. This is excluded by the assumption the
variational principle is proper. We have thus proven that

\medskip\noindent
{\bf Proposition 12.} {\it The characteristic vector field $W$ for the variational principle defined by $\vth$ satisfies $W \interno \d \vth = 0$, i.e. $W \in \N (\d \vth)$.} \EOS
\medskip

Recalling that $\N (\d \vth)$ is one dimensional (see also remark
1 for its explicit description) we have in fact shown that $\N (\d
\vth )$ coincides with the one-dimensional module generated by the
characteristic vector field $W$ for the variational principle
identified by $\vth$.

We can summarize our discussion by introducing a suitable definition:

\medskip\noindent
{\bf Definition 7.} {\it A vector field $W$ on $M$ satisfying $W \interno \d \vth = 0$, i.e. $W \in \N (\d \vth)$ is a {\bf  characteristic vector field} for the maximal degree proper variational principle on $\pi : M \to B$ defined by $\vth$.} \EOD
\medskip

\subsection{Liouville dynamics}

A vector field in the phase space $P$ is said to be {\it Liouville}
-- or to define a Liouville dynamics -- if it preserves a volume in
phase space. The geometry of Liouville vector fields has been
discussed by several authors in parallel with the geometry of
Hamilton vector field, see e.g. \cite{Mar}. Here we show how our
discussion for maximal degree variational principles applies to
Liouville dynamics; see \cite{GMhh,GMlio,GMvar} for further detail.

Note in this case there is a preferred independent variable, i.e. time. We will thus write $B = \R \times Q$ and $M = \R \times P$. We assume $P$ is a connected orientable manifold, and denote by $\Om$ the reference volume form on it. We will choose a form $\s$ such that $\d \s = \Om$.

The vector field $X$ on the phase space $P$ is (globally) Liouville with respect to $\Om$ if there is a form $\ga$ such that
$$ X \interno \Om \ = \ \d \ga \ . \eqno(5) $$

To $X$ we associate a vector field $Z = \pa_t + X$ on the extended
phase space $M = \R \times P$. One can then prove the following
result.

\medskip\noindent
{\bf Proposition 13.} {\it Let $X$ be a Liouville vector field on $P$, $X \interno \Om = \d \ga$ and $Z = \pa_t + X$ be the associated vector field on $M = \R \times P$. Then $Z$ is the unique characteristic vector field for the maximal degree variational principle on $\pi : M \to B$ defined by
$$ \vth \ := \s \, + \, (-1)^s \, \ga \w \d t \ \in \ \Lambda^{n-2} (M)  \eqno(6) $$
(with $s = \pm 1$ depending on orientation) satisfying $Z \interno \d t = 1$.} \EOS
\medskip

It may be worth mentioning that $\vth$ can be determined via Hodge duality \cite{GMlio}. Denote as usual by $* \a$ the Hodge dual to the form $\a$; and by $(\widetilde{Z})$ the one-form in $M$ dual to the vector field $Z$: if $Z = \pa_t + f^i \pa_i$, this will be $\d t + g_{ij} f^j \d x^i$, with $g$ the metric in $P$. Then we have:

\medskip\noindent
{\bf Proposition 14.} {\it The form $\vth$ defining the variational principle associated to the Liouville vector field $X$ satisfies $ \d \vth \ = \ \sqrt{|g^{-1}|} \ *(\widetilde{Z})$.} \EOS
\medskip

Note that this condition completely determines the variational principle: indeed it identifies $\vth$ up to a closed form, which has no role in the variation of $I(\phi) = \int_D \phi^* (\vth )$.

\section{The decomposable case}

As discussed in section 4, the reduction procedure based on
proposition 4 is in general not viable, as $\N (\d \vth )$ fails
to exist. There {\it are}, however, cases in which the reduction
procedure discussed above can be performed.

We will deal with the simplest occurrence of the case, i.e. that
where $\d \vth$ is a decomposable form. This is enough to show the
main ingredient of the procedure discussed in sect.4 at work, and
to describe a multidimensional module of vector fields in terms of
a variational principle. A more general discussion will be given elsewhere.

\subsection{The decomposable case in general}

In particular, let us consider the case where $\eta = \d \vth$ is a
{\bf decomposable} form, i.e. there are 1-forms $\a_i$
($i=1,...,k+1$) such that $$ \eta \ := \ \d \vth \ = \ \a_1 \w ... \w
\a_{k+1} \ . \eqno(7) $$ In this case we will say that $\vth$ is
{\it $\d$-decomposable}.

We will denote by $\N_\pi (\eta )$ the subset of vector fields in
$\N (\eta )$ which are transversal to fibers of $\pi : M \to B$.

\medskip\noindent
{\bf Definition 8.} The decomposable form $\eta \in \La^{k+1} (M)$
is {\bf nondegenerate} if the forms $\{ \a_i \}$ are independent
at all points $x \in M$. It is {\bf compatible} with the fibration
$\pi$ if the dimension of $\N_\pi (\eta)$ is constant, and {\bf
adapted} to $\pi : M \to B$ if $\N (\eta ) \cap \V (\pi) =
\emptyset$, i.e. if $\N_\pi (\eta) = \N (\eta)$. \EOD

\medskip\noindent
{\bf Lemma 2.} {\it Let $M$ be a $n$-dimensional manifold, and
$\eta \in \La^{k+1} (M)$ a nondegenerate decomposable form. Then
$\N (\eta)$ is a $q = (n-k-1)$ dimensional module over $\La^0
(M)$.} \EOS

\medskip\noindent
{\bf Proof.} As $\eta$ is decomposable and nondegenerate, we have
$$ \N (\eta) \ = \ \N (\a_1 ) \cap ... \cap \N (\a_{k+1} ) \ ;  $$
note that each $\N (\a_i )$ spans a distribution of codimension
one, hence $\N ( \eta) $ has codimension $k+1$, i.e. dimension
$n-k-1$.

Equivalently, denote by $Y_i$ the vector field dual to the
one-form $\a_i$; the nondegeneration of $\d \eta$ implies that the
$\{ Y_1 , ... , Y_{k+1} \}$ span at each point $x \in M$ a
$(k+1)$-dimensional subspace ${\cal Y}_x$ of $T_x M$. The vector
fields $X \in \N (\d \vth )$ are then vector fields which are in
the orthogonal complement to ${\cal Y}_x$ at each point $x \in M$,
hence they span a $(n-k-1)$-dimensional module. \EOP
\medskip

In this case, not only $\N (\d \vth)$ is not empty (see sect.4),
but has dimension $q=(n-k-1)$. Note that $\vth$ is of degree $k$,
and $B$ of dimension $k$; this means that critical sections will
be submanifolds of $M$ also of dimension $k$. Thus, in order to
have a proper variational principle based on a form $\vth$ such
that $\eta := \d \vth$ is decomposable and nondegenerate, a
necessary (but not sufficient) condition is that $n-k-1 \le k$,
i.e.
$$ n \ \le \ 2 k +1 \ . \eqno(8) $$
We will refer to the case $n = 2k+1$ as the {\bf maximally
characteristic} case.

We also recall that for the $\d$-decomposable form $\vth$ to
define a proper variational principle in $(M,\pi,B)$, it is
necessary that none of the vectors in $\N (\d \vth)$ is vertical,
i.e. that $\d \vth$ is adapted to $\pi : M \to B$.

\medskip\noindent
{\bf Lemma 3.} {\it Let $\vth \in \La^k (M)$ be a
$\d$-decomposable form, such that it defines a proper variational
principle in $(M,\pi,B)$ and $\d \vth = \a_1 \w ... \w \a_{k+1}$.
Then the complete ideal $\=\J (\vth,\pi)$ associated to the
Cartan variational ideal is generated by $\{ \a_1 , ... , \a_{k+1}
\}$.} \EOS

\medskip\noindent
{\bf Proof.} As the $\a_i$ are independent, a vector field $Y$ in
$M$ can satisfy $Y \interno \eta = 0$ if and only if $Y \interno
\a_i = 0$ (this, of course, is again the remark that $\N
(\eta ) = \N (\a_1 ) \cap ... \cap \N (\a_{k+1} )$). From this
and the definition of $\=\J (\vth,\pi)$ the statement is
immediate. \EOP

\medskip\noindent
{\bf Lemma 4.} {\it If $\vth$ is such that $\d \vth$ is
decomposable, nondegenerate and compatible with the fibration
$\pi$, then $\D = D[\J (\pi,\vth)]$ is integrable.} \EOS

\medskip\noindent
{\bf Proof.} The properties assumed on $\d \vth = \eta$ imply that the complete ideal $\bar{\J} (\vth , \pi)$ is a differential ideal. In fact, $\bar{\J} (\vth , \pi )$ is generated by the $\a_i$, and $\d \eta = 0$ guarantees that $\d \a_i \in \bar{\J} (\vth , \pi )$ as well. The characteristic distribution $\D$ of $\J (\vth , \pi)$ is also the characteristic distribution of $\bar{\J} (\vth , \pi)$, by definition, and proposition 2 implies this is integrable. \EOP
\medskip

It will be convenient to state some simple general results, also
to establish a convenient notation for our later discussion. We
stress that here we assume the condition (8) is satisfied; see
sect.7 for the opposite case.

We work in a local chart, i.e. in $\R^n$ (recall that we deal with
a local variational principle), and we will deal with the case of
euclidean metric.\footnote{The modifications needed to take a more
general metric into account are rather obvious, but dealing with
this case will keep notation simpler.}

We introduce local orthogonal coordinates $\{ x^1 , ... , x^k \}$
in $B$, and $\{ z^1 , ... , z^p \}$ (with $p = n-k \le k+1$ for
the variational principle to be proper; and necessarily $n \ge k+2$, see footnote 3) on the fiber $F \simeq \pi^{-1} (x)$. We write $h = 2k +1 - n$, hence $p = k+1 - h$ and (8) implies $0 \le h \le k-1$.

It is convenient to write the forms $\a_i$ as
$$ \a_i \ = \ M_{ij} \d x^j \ + \ L_{ia} \d z^a $$
where $i=1,...,k+1$ and the dummy indices $j$ and $a$ run,
respectively, from 1 to $k$ and from 1 to $p$. The $(k+1) \times
k$ dimensional matrix $M$ and the $(k +1) \times p$ dimensional
matrix $L$ are of course functions of $(x,z)$.

As we assumed $\eta$ to be nondegenerate and adapted to the
fibration $\pi$, the rank of $L$ is constant and equal to $p <
k+1$. We can thus, with a point-dependent change of coordinates
(or considering linear combinations of the $\a_i$), take a square
submatrix $L_0$ of $L$ -- say the one given by its first $p$ rows
-- to diagonal form and set to zero the remaining rows; with a
rescaling of the (new) $z$ coordinates, $L_0$ can be assumed to be
the identity. In this way, we can limit to deal with
$$ \cases{ \a_j \ = \ \d z^j + B_{jm} \d x^m & for $j = 1,...,p$; \cr
\a_{p+i} \ = \ C_{im} \d x^m & for $i = 1,...,h$ \cr} \eqno(9) $$
(no confusion should be possible between the matrix $B$ and the
base manifold of $\pi:M \to B$). Note now that
$$ {\pa \over \pa z^a} \interno \a_j \ = \ \cases{ \de_{aj} & for $j \le p$, \cr
0 & for $j > p$. \cr} \eqno(10) $$

We introduce now the decomposable forms $\chi_s \in \Lambda^k (M)$
obtained by the wedge product of all the $\a_i$ but $\a_s$, with a
factor $(-1)^{s-1}$. That is,
$$ \chi_s \ := \ (-1)^{s-1} \ \a_1 \w ... \w \a_{s-1} \w \a_{s+1}
 \w ... \w \a_{k+1} \ . \eqno(11) $$

\medskip\noindent
{\bf Lemma 5.} {\it With $\a_i$ as in (9) and $\eta$ given by (7),
we have
$$ \Psi_a \ := \ {\pa \over \pa z^a} \interno \eta \ = \
\chi_a  \eqno(12) $$} \EOS

\medskip\noindent
{\bf Proof.} This follows immediately from (10) and (11). \EOP

\medskip\noindent
{\bf Theorem 1.} {\it Let $\vth$ define a proper variational
principle in $(M,\pi,B)$. Then, with the notation introduced above
where $\eta := \d \vth$, the equations $\De_a = 0$ identifying
critical sections $\phi \in \Ga (\pi)$ for the variational
principle defined by $\vth$ are given by
$$ \phi^* (\Psi_a ) \ = \ \phi^* (\chi_a ) \ = \ 0 \ . \eqno(13) $$} \EOS

\medskip\noindent
{\bf Proof.} This is a restatement of our discussion, and follows
immediately from Lemma 5. \EOP
\medskip

The forms $\phi^* (\Psi_a) \in \La^k (B)$ can necessarily be
written as $\De_a \om$, where $\om := \d x^1 \w ... \w \d x^k$ is
the volume form in $B$. Thus (13) can also be written as
$$ \De_a \ = \ 0 \ ; \eqno(14) $$
these are first order nonlinear PDEs for the dependent variables
$z^a$ in terms of the independent variables $x^i$.

We will now provide a compact way of writing the $\De_a$. Given a
matrix $P$ of dimension $(k+1)\times k$, we will denote by $\^P_i$
the $k \times k$ matrix obtained suppressing the $i$-th row, and by  $||\^P_{i}||$ its determinant with a factor $(-1)^{i+1}$.

We recall that if we have $k$ one-forms $\b_i = A_{ij} \d x^j \in
\La^1 (\R^k )$, and write $\om = \d x^1 \w ... \w \d x^k$, then
$$ \b_1 \w ... \w \b_k \ = \ ||A|| \ \om \ . \eqno(15) $$

\medskip\noindent
{\bf Theorem 2.} {\it Consider a variational principle on
$(M,\pi,B)$ identified by $\vth$ such that $\d \vth$ is
decomposable, nondegenerate and compatible with the fibration
$\pi$. Write $\eta = \d \vth $ as in (7) and in general use the
notations introduced in this section. Then the equations (1)
identifying critical sections are written as $||\^P_{a} || = 0 $
for $a=1,...,p$\footnote{Note no condition is set on $||\^P_j ||$ for $j>p$.}, with the $(k+1)\times k$ matrix $P$ given by
$$ P_{ij} \ := \ \cases{B_{ij} \, + \, {\pa z^i \over \pa x^j } &
for $i=1,...,p$, \cr C_{i-p, j} & for $i=p+1,...,h$. \cr}
\eqno(16)
$$} \EOS

\medskip\noindent
{\bf Proof.} With $\a_i$ in the form (9), we have
$$ \cases{\phi^* (\a_i ) \ = \ [ B_{im} + (\pa z^i / \pa x^m ) ] \, \d x^m  &
for $i \le p$, \cr \phi^* (\a_{p+j} ) \ = \  C_{jm} \, \d x^m &
for $j \le h$. \cr} $$ Note that, with an abuse of notation, we
write $f$ for $\phi^* (f)$ when dealing with functions.

It follows from the definition (11) of $\chi_s$, together with
(12), (15) and standard properties of the pullback operation, that
$$ \begin{array}{rl}
\phi^* (\Psi_a ) \ = & \[ {1 \over p!} \, \epsilon_{a i_1 ... i_p}
\, P_{i_1 j_1} ... P_{i_p j_p} \] \, C_{\ell_1 j_{p+1}} ...
C_{\ell_h j_k} \  \d x^{j_1} \w ... \w \d x^{j_k} \ = \\
= &  {1 \over k! p!} \, \epsilon_{i j_1 ... j_k} \ \epsilon_{a i_1
... i_p} \, P_{i_1 j_1} ... P_{i_p j_p} \, C_{\ell_1 j_{p+1}} ...
C_{\ell_h j_k} \, \om \ := \ \De_i \, \om \end{array} \eqno(17) $$
with $P$ given indeed by (16). It suffices now to note that
$$ \epsilon_{i j_1 ... j_k} \ \epsilon_{a i_1 ... i_p} \, P_{i_1
j_1} ... P_{i_p j_p} \, C_{\ell_1 j_{p+1}} ... C_{\ell_h j_k} $$
coincides with $|| \^P_i ||$. \EOP

\medskip\noindent
{\bf Remark 6.} For the maximally characteristic case $n = 2k +1$,
we have $h = 0$ and one should understand $C = 0$ in the above
discussion and formulas; see next subsection. \EOR
\medskip

The expression (17) also provides a way of writing the equations
$\De_a = 0$ in terms of a certain set of vector fields in $M$.
Indeed, introducing the vector fields
$$ X_i \ := \ {\pa \over \pa x^i } \ + \ B_{ai} \, {\pa \over \pa
z^a} \ \ \ \ (i=1,...,k) \ , $$
we rewrite (16) as
$$ P_{ij} \ := \ \cases{ X_j (z^i) & for $i=1,...,p$, \cr
C_{i-p, j} & for $i=p+1,...,h$. \cr} $$

This makes clear that the equations $\De_a = 0$ will be written in
the form (we omit a combinatorial factor $[k! (p-1)!]^{-1}$)
$$ \De_a \ := \ \epsilon_{m_1 ... m_k} \, \epsilon_{a b_1 ... b_{p-1} } \, \[ X_{m_1} (z^{b_1}) ...
X_{m_{p-1}} (z^{b_k}) \] \ \Theta_{m_p ... m_k} \ = \ 0 \ , $$ where
$\Theta$ depends only on $x$ and $z$, not on the derivatives $\pa_i z^j$.

\subsection{The maximally characteristic case}

Let us now consider the case $n = 2k+1$; as we have seen before,
in this case ${\rm dim} [\N (\d \vth)] = k$. If $\N (\d \vth)$ is
nowhere vertical, i.e. if $\d \vth$ is adapted to the fibration
$\pi : M \to B$ and hence $\vth$ defines a proper variational
principle, then the critical sections will be spanned by the
distribution $\N (\d \vth )$.

Thus the (partial differential) equations issued by the
variational principle should be equivalent to a set of (ordinary
differential) equations, each of them defining a vector field on
$M$; these vector fields in turn generate the module $\N (\d
\vth)$\footnote{Note that this remark is not interesting for the
maximal degree case $n-k = 2$: indeed $k+2 = n = 2k + 1$ enforces
$k=1$ (and $n=3$), i.e. we would be in a case where the
variational principle is defined by a one-form, and is thus
obvious it produces a vector field -- or more precisely a
direction field, see remark 1 above.}.
This is indeed what happens. We will make it precise in the
following statement.

\medskip\noindent
{\bf Theorem 3.} {\it Let $\pi : M \to B$ be a smooth fiber bundle of
dimension $n=2k+1$ with base manifold $B$ of dimension $k$; let $\vth \in \Lambda^k (M)$ such that $\d \vth$ is decomposable, nondegenerate and adapted to the fibration $\pi$. Then $\vth$ defines a proper variational principle in $(M,\pi,B)$, and the variational Cartan ideal $\J (\vth,\pi)$ is nonsingular and admits a $k$-dimensional
characteristic distribution $\D = D[\J (\vth,\pi )]$; this coincides with $\N (\d \vth )$. Moreover  $\D$ is completely
integrable and integral manifolds of $\D$ coincide with critical
sections for the variational problem defined by $\vth$.} \EOS

\medskip\noindent
{\bf Proof.} First of all we note that now $q = n - (k-1) = k$,
hence the dimension of $\D$ given in the statement agrees with our
general results.

We have seen that a section $\varphi\in \Gamma (\pi)$ is critical
for the variational problem defined by $\vth$ if and only if
$\varphi$ is an integral manifold for the variational ideal $ \J
(\vth,\pi )$ (see proposition 7). Moreover, in the decomposable
case, the characteristic distribution $D[\J (\vth,\pi )] \equiv
\D$ of the variational ideal $\J (\vth,\pi )$ is a completely
integrable $k$-dimensional distribution, see lemma 4.

Then, recalling that integral manifolds of $\D$ are also integral
manifolds of $\J (\vth,\pi)$, we can conclude, using also the
compatibility condition between $\vth$ and the fibration $\pi : M
\to B$, that critical sections for the variational principle can be
identified with integral manifolds of the distribution  $D[\J
(\vth,\pi )]$. \EOP

\medskip\noindent
{\bf Remark 7.} Note that the condition $||\^P_i || = 0 $ for all
$i=1,... , p$ means that ${\rm rank} (P) < k$. \EOR
\medskip

In order to help comparison with later results, we give a
corollary which is essentially a restatement of theorem 3:

\medskip\noindent
{\bf Corollary 1.} {\it In the maximally characteristic case, a
variational principle in the bundle $(M,\pi,B)$ based on $\vth \in
\La^k (M)$ satisfying the hypotheses of theorem 3, uniquely
identifies the $k$-dimensional integrable distribution $\D = D[\J
(\vth , \pi )]$; this coincides with the module $\N (\d \vth)$ of
vector fields which are tangent to all the critical sections, and
conversely critical sections are the manifolds for which all
tangent vectors are in $\N (\d \vth )$.} \EOS

\subsection{The non maximally characteristic case}

In the more general case $n < 2k +1$, we will have a situation
similar to that described by Proposition 9. We will reformulate
the latter for the case at hand.

Recall preliminarily that $\N (\d \vth )$ is $q$-dimensional, with
$q = n-k-1$ (see lemma 2), and $n = 2k +1 -h$, with $0 \le h \le
k-1$;  thus we also have $q = k-h$. The extremal cases $h=k-1$ and
$h=0$ correspond, respectively, to the case of maximal degree
variational principles and to the maximally characteristic case.
Here we consider the case $0 < h < k-1 $.

\medskip\noindent
{\bf Theorem 4.}  {\it Let the $\d$-decomposable form $\vth$
define a proper variational principle in $(M , \pi , B)$. Let $B_0
\ss B$ be a smooth $h$-dimensional submanifold of $B$, and $\pi_0
: \pi^{-1}(B_0) \to B_0$ the associated subbundle of $\pi: M \to
B$. Let the section $\phi_0 \in \Ga (\pi_0)$ be an integral
manifold for the Cartan ideal $\J (\vth,\pi)$, nowhere tangent to
integral manifolds of $D[\J (\vth,\pi)]$. Then the critical local
sections for the variational principle defined by $\vth$ can be
built by pulling $\phi_0$ along local integral manifolds of the $q
= k-h$ dimensional distribution $D [\J (\vth,\pi)]$.} \EOS
\medskip

\medskip\noindent
{\bf Proof.} From lemma 4 we know that the characteristic distribution  $D[\J (\vth,\pi)]$ of the variational ideal is completely integrable and $q$-dimensional, with $q = k-h$.
Moreover, by proposition 4, we can build local integral manifolds
of the differential ideal $\J (\vth,\pi)$ by pulling a lower
dimensional local integral manifold along the local integral
manifold of the characteristic distribution $D[\J (\vth,\pi)]$ (as
all vector in $D[\J (\vth,\pi)]$ are tangent to integral manifold
of $\J (\vth,\pi)$).

Then, let us start from the submanifold $\phi_0 \ss M$: this is an
integral manifold for the Cartan ideal $\J (\vth,\pi)$, nowhere
tangent to integral manifolds of $D[\J (\vth,\pi)]$. Hence we can
obtain a local integral manifolds $\Phi$ for $\J (\vth,\pi) $,
such that $\phi_0 \ss \Phi$, pulling along local integral
manifolds of $\D$, see proposition 4.

Recalling that $\vth$ is nondegenerate and adapted to the
fibration, and proposition 7, we conclude that the $k$-dimensional
submanifolds obtained in this way are also critical sections for
the variational principle defined by $\vth$. \EOP
\medskip

In this case we will also say that vector fields in $\N (\d \vth)$
are characteristic vector fields for the variational principle
identified by $\vth$, see definition 6.

\medskip\noindent
{\bf Corollary 2.} {\it In the non maximally characteristic case,
a (proper) variational principle in the $n$-dimensional fiber
bundle  $(M,\pi,B)$ based on $\vth \in \La^k (M)$ (with $k+2 < n <
2k +1$) such that $\vth$ is $\d$-decomposable, nondegenerate and
adapted to the fibration $\pi$, uniquely identifies the
$q$-dimensional ($q = n-k-1 < k$) integrable distribution $\D =
D[\J (\vth , \pi )]$; this coincides with the module $\N (\d
\vth)$ of vector fields which are tangent to all the critical
sections.} \EOS

\section{Non proper variational principles}

In this section we want to discuss the case where $\d \vth$ is
decomposable, nondegenerate and compatible with the fibration $\pi
: M \to B$, but not adapted to it. That is, $\N (\d \vth )$ will
include some vertical vector field.

Note that as $\d \vth$ is nondegenerate, $\D = D [ \J (\vth , \pi
) ]$ is a distribution, i.e. $\N (\d \vth)$ has constant rank.
This, in combination with the assumption $\d \vth$ is compatible
with $\pi$ (recall this means that $\D_\pi$ has constant
dimension) implies that the module $\N ( \d \vth ) \cap \V (\pi )$
has also constant dimension, as it is the complementary to
$\D_\pi$ in $\N (\d \vth )$.

In this case we introduce local orthogonal coordinates $\{ x^1 ,
... , x^k \}$ in $B$, and $\{ y^1 , ... , y^p \}$ (with $p = n-k
\ge k+1$) on the fiber $F \simeq \pi^{-1} (x)$.

It will be convenient to separate the vertical coordinates in two
subsets, i.e. a set of $k+1$ ones, which we denote as $\{ z^1 ,
... , z^{k+1} \}$, and a residual set of $s = p - (k+1) > 0$ ones
which we denote as $\{ w^ 1 , ... , w^s \}$. The reason for this
splitting is the following.

We can write, in full generality, the forms $\a_i$ as
$$ \a_i \ = \ L_{ia} \d y^a  \ + \  M_{ij} \d x^j \ ; $$
here $L$ is a $(k+1) \times p$ dimensional matrix, and $M$ is a
$(k+1) \times k$ dimensional one (both of these are a function of
the point $(x,y)$, of course). Assuming that $\eta$ is
nondegenerate and compatible with the fibration $\pi$, the rank of
the matrix $L$ is constant and equal to $k+1 < p$; thus there
exists a change of coordinates (depending on $\xi$) in which a
$(k+1)$-dimensional square submatrix of $L$ is diagonal (see the
discussion in sect.6). The $z^a$ will be the corresponding
coordinates. Thus we write
$$ \a_i \ = \ \d z^i \ + \ B_{ij} \d x^j \ + \ G_{im} \d w^m \ . \eqno(18) $$

We introduce now the decomposable forms $\chi_s \in \Lambda^k (M)$
defined as in sect.6, i.e. obtained by the wedge product of all
the $\a_i$ but $\a_s$, with a factor $(-1)^{s-1}$.

\medskip\noindent
{\bf Lemma 6.} {\it With $\eta$ given by (7), and $\a_i$ as in (18), we have
$$ {\pa \over \pa z^a} \, \interno \, \eta \ = \ \chi_a \ \ , \ \
{\pa \over \pa w^m} \, \interno \, \eta \ = \ G_{jm} \chi_j \ .
\eqno(19) $$}

\medskip\noindent
{\bf Proof.} The first equation is just the definition of
$\chi_s$, due to (18). The second follows immediately from the
expression of the $\a_i$ and of $\eta$. \EOP
\medskip

Consider the variational principle on $(M,\pi,B)$, defined by
$\vth$, and the associated variational ideal $\J (\pi,\vth)$. This
is generated by $\{ \Psi_j \}$ with $j=1,...,p$. We can decide in
full generality that
$$ \Psi_a \ = \ \cases{ (\pa / \pa z^a) \interno \d \vth & for
$a = 1,..., k+1$ \cr (\pa / \pa w^m) \interno \d \vth & for $a =
k+1+m, \ m=1,..., s$ \ . \cr} \eqno(20) $$

\medskip\noindent
{\bf Lemma 7.} {\it If $\vth$ is such that $\d \vth$ is
decomposable, nondegenerate and compatible with the fibration
$\pi$, then with the choice (20), $\J (\pi,\vth)$ is generated by
$\{\Psi_1 , ... , \Psi_{k+1} \}$.} \EOS

\medskip\noindent
{\bf Proof.} It follows from (19) and (20) that
$$ \Psi_{k+1+m} \ = \ \sum_{a=1}^{k+1} \ G^T_{ma} \ \Psi_a \ . \eqno(21) $$
The lemma is an immediate consequence of (21). \EOP
\medskip

The equations identifying critical sections $\phi$ given in
coordinates by $y^a = y^a (x)$ will be obtained simply from the
requirement
$$ \De_a \ := \ \phi^* (\Psi_a) \ = \ 0 \ \ \ {\rm for} \ a=1,...,k+1 \ . \eqno(22) $$
Indeed, if these are satisfied, the ones for $a=k+1+m$,
$m=1,...,s$ are also satisfied, see (21). Needless to say, this is
just another way of seeing lemma 7.

\bigskip

We will now provide a compact way of writing the equations $\De_a
= 0 $ ($a = 1,...,k+1$) identifying critical sections, similarly
to what was done in sect.6 and using the same notation. That is,
given a matrix $P$ of dimension $(k+1)\times k$, we will denote by
$||\^P_{i}||$ the determinant of the $k \times k$ matrix obtained
suppressing the $i$-th row (with a sign $(-1)^{i-1}$).

\medskip\noindent
{\bf Lemma 8.} {\it Consider a variational principle on
$(M,\pi,B)$ identified by $\vth$ such that $\d \vth$ is
decomposable, nondegenerate and compatible with the fibration
$\pi$. Write $\eta = \d \vth $ as in (7), with $\a_i$ as in (18).
Then the equations (22) identifying critical sections are written as $||\^P_{a} || = 0 $ (for all $a=1,...,k$) for $P$ the $(k+1)\times k$ matrix given by
$$ P_{ij} \ = \ B_{ij} + (\pa z^i / \pa x^j) + G_{im} (\pa w^m /
\pa x^j) \ , $$ where $B$ and $G$ are the matrices appearing in (18).} \EOS

\medskip\noindent
{\bf Proof.} With $\a_i$ given by (18), we have
$$ \phi^* (\a_i) \ = \ \[ B_{ij} \, + \, {\pa z^i \over \pa x^j} \, + \,
G_{im} {\pa w^m \over \pa x^j} \] \ \d x^j \ := P_{ij} \d x^j \ .
\eqno(23) $$ Proceeding as in the proof of theorem 2, we identify
the condition of vanishing of $\phi^* (\Psi_a) $ for $a = 1,...,p$
with the condition that $||\^P_a || = 0 $, or equivalently ${\rm rank} (P) < k$. \EOP
\medskip

We can also rewrite $P$ in terms of the action of vector fields,
similarly to the case of proper variational principles. We define
$$ X_i \ := \ {\pa \over \pa x^i} \ + \ B_{ji} {\pa \over \pa z^j}
\eqno(24) $$ and with these we have
$$ P_{ij} \ = \ X_j (z^i) \ + \ G_{im} \, X_j (w^m) \ . \eqno(25) $$

Let us now come to the reduction theorem in this case, i.e. the analogue of theorems 3 and 4. We preliminarily note that the first part of these theorems, ensuring the variational principles identifies an integrable distribution -- which is just the module $\N (\d \vth )$ -- of vector fields, has a counterpart in this case:

\medskip\noindent
{\bf Theorem 5.} {\it Let $\pi : M \to B$ be a smooth fiber bundle
of dimension $n=2k+1+s$ ($s>0$) with base manifold $B$ of
dimension $k$; let $\vth \in \Lambda^k (M)$ such that $\d \vth$ is
decomposable, nondegenerate and compatible with the fibration
$\pi$. Then $\vth$ defines a (necessarily non proper) variational
principle in $(M,\pi,B)$; the variational Cartan ideal $\J
(\vth,\pi)$ is nonsingular and admits a $(k+s)$-dimensional
characteristic distribution $\D = D[\J (\vth,\pi )]$; this
coincides with $\N (\d \vth )$. Moreover  $\D$ is completely
integrable.}

\medskip\noindent
{\bf Proof.} First of all we note that now $q = n - (k-1) = k+s$,
hence the dimension of $\D$ given in the statement agrees with our
general results.

Proceeding as in the proof of theorem 3, we recall that in the
decomposable case, the characteristic distribution $\D$ of the
variational ideal $\J (\vth,\pi )$ is a completely integrable
$(k+s)$-dimensional distribution, see lemma 4. \EOP
\medskip

Let us now discuss how $\D = \N (\d \vth )$ can be used for
reduction in the spirit of section 4 in this case. From
proposition 2 we have at once that integral manifolds of $\D$ are
also integral manifolds for the variational ideal $\J = \J (\vth ,
\pi)$. Proposition 4 would also allow to build integral manifolds
of $\J$ starting from non-characteristic lower dimensional
integral manifolds and pulling them along integral manifolds of
$\D$.

Note however that we are not interested in generic integral
manifolds for the variational ideal $\J (\vth , \pi)$, but only in
those which are also sections for the bundle $\pi : M \to B$ (i.e.
critical sections). Thus, roughly speaking, we should use only the
part of $\D$ which is transversal to fibers $\pi^{-1} (b)$, i.e.
$\D_\pi$, for pulling lower dimensional integral manifolds of
$\J$. Note that $\D_\pi$ has constant dimension $r$ since $\vth$
is compatible with the fibration $\pi$; we assume that $r > 0$.

The problem with using $\D_\pi$ to generate higher dimensional
integral manifolds lies in that proposition 4 relies on the fact
that $\D$ is integrable; but integrability of $\D$ does not imply
integrability of $\D_\pi$, as the commutator of transversal vector
fields could fail to be transversal. Hence we are not guaranteed
$\D_\pi$ is an integrable distribution and in general we can not
just use this for our reduction procedure (see the example in
sect.10 for an illustration of this).

In the very special case where $\D_\pi$ is integrable, we can
state a very close analogue of theorems 3 and 4:

\medskip\noindent
{\bf Lemma 9.} {\it Let the hypotheses of theorem 5 be verified.
Assume moreover that $\D_\pi \ss \D$ is an integrable
distribution. If  $\D_\pi$ has dimension $k$, then critical
sections for the variational problem defined by $\vth$ coincide
with integral manifolds for $\D_\pi$.} \EOS

\medskip\noindent
{\bf Proof.} Follow the proof of theorem 3, using $\D_\pi$ rather
than $\D$. \EOP

\medskip\noindent
{\bf Lemma 10.} {\it Let the hypotheses of theorem 5 be verified.
Assume moreover that $\D_\pi \ss \D$ is an integrable
distribution, of dimension $r=k-h$ ($0<h<k$). Let $B_0 \ss B$ be a
smooth $h$-dimensional submanifold of $B$, $\pi_0 : \pi^{-1}(B_0)
\to B_0$ the associated subbundle of $\pi: M \to B$, and $\phi_0
\in \Ga (\pi_0)$ be an integral manifold for the Cartan ideal $\J
(\vth,\pi)$, nowhere tangent to integral manifolds of $\D_\pi$.
Then the critical local sections for the variational principle
defined by $\vth$ can be built by pulling $\phi_0$ along integral
manifolds of the $r = k-h$ dimensional distribution $\D_\pi$.}
\EOS

\medskip\noindent
{\bf Proof.} Follow the proof of theorem 4, using $\D_\pi$ rather than $\D$. \EOP
\medskip

Let us now consider the general case, i.e. the one where we are
not guaranteed that the $r$-dimensional distribution $\D_\pi$ is
integrable. We write again $r = k-h$ with $0 \le h < k$; for
$h=0$, the role of $B_0$ and $\phi_0$ in the theorem below is
played by any point $m_0 \in \pi^{-1}(b_0)$.

\medskip\noindent
{\bf Theorem 6.} {\it Let the hypotheses of theorem 5 hold, and
let $\D_\pi$ be of dimension $k$. Then critical sections for the
variational problem defined by $\vth$ are submanifolds of integral
manifolds of $\D$, and their tangent vector fields belong to
$\D_\pi$.} \EOS

\medskip\noindent
{\bf Proof.} Integral manifolds of $\D$ are also integral
manifolds of $\J (\vth,\pi)$; thus submanifolds of the former are
submanifolds of the latter. This applies in particular to
submanifolds   of integral manifolds of $\D$ whose tangent vectors
are in $\D_\pi$; such submanifolds are $k$-dimensional and
everywhere transversal to fibers of $\pi : M \to B$ and are thus
sections of this bundle. This also implies they are critical
sections for the variational principle defined by $\vth$.
Conversely, consider a critical section $\phi$ through a point
$m$: this necessarily belongs to the integral submanifold through
$m$ of the integrable distribution $\D$, and tangent vectors to
$\phi \ss M$ are necessarily in $\D_\pi \ss \D$. \EOP

\medskip\noindent
{\bf Theorem 7.}  {\it Let the hypotheses of theorem 5 hold, and
let $\D_\pi$ be of dimension $r=k-h$, $0< h < k$. Let $B_0 \ss B$
be a smooth $h$-dimensional submanifold of $B$, and $\pi_0 :
\pi^{-1}(B_0) \to B_0$ the associated subbundle of $\pi: M \to B$.
Let the section $\phi_0 \in \Ga (\pi_0)$ be an integral manifold
for the Cartan ideal $\J (\vth,\pi)$, nowhere tangent to integral
manifolds of $\D = D[\J (\vth,\pi)]$. Then the local critical
sections $\phi$ for the variational principle defined by $\vth$
are submanifolds of the manifolds built by pulling $\phi_0$ along
integral manifolds of the $q = k+s$ dimensional distribution $\D$;
the tangent vector fields to $\phi$ are in $\D_\pi$.} \EOS
\medskip

\medskip\noindent
{\bf Proof.} The distribution $\D$ is integrable (proposition 2).
Thus, by proposition 4, we can build a local integral manifold
$\Phi$ of the differential ideal $\J (\vth,\pi)$ by pulling
$\phi_0 \ss M$ along the local integral manifold of $\D$. Note
that as $\D_\pi$ has dimension $r = k-h$ and $\phi$ has dimension
$h$, for any point $m \in \Phi \ss M$ there is a subspace of $\T_m
\Phi$ which is transversal to fibers of $\pi$ and of dimension
$k$. This means that there are $k$-dimensional submanifolds $\phi
\ss \Phi$ which are transversal to fibers of $\pi$, i.e. which are
sections for $\pi : M \to B$. As they are also integral manifolds
for $\J (\vth , \pi )$, they are indeed critical sections for the
variational principle defined by $\vth$. The tangent vector fields
to these are by construction in $\D$, and transversality ensures
they are actually in $\D_\pi$. Conversely, consider a critical
section $\^\phi$ such that $\phi_0 \ss \^\phi$: necessarily this
is a submanifold of $\Phi$ considered above, $\^\phi \ss \Phi$,
and by unicity we conclude that actually $\^\phi = \phi$. \EOP

\medskip\noindent
{\bf Corollary 3.} {\it The (necessarily non proper) variational
principle in the $n$-dimensional fiber bundle  $(M,\pi,B)$ based
on $\vth \in \La^k (M)$ (with $n > 2k +1$) such that $\vth$ is
$\d$-decomposable, nondegenerate and compatible with the fibration
$\pi$, uniquely identifies the integrable distribution $\D = D[\J
(\vth , \pi )]$, which coincides with the module $\N (\d \vth)$.
If the transverse part $\D_\pi$ of $\D$ have positive dimension,
then vector fields in $\D_\pi$ -- i.e. non vertical vector fields
in $\N (\d \vth )$ -- are tangent to all critical sections for the
variational principle.} \EOS

\section{Example 1: maximally characteristic case}

Let us consider the space $M = \R^n$, seen as a fiber bundle
$(M,\pi,B)$ on the space $B = \R^k$, and a $\d$-decomposable form
$\vth \in \La^k (M)$ defining a proper variational principle in
$(M,\pi,B)$; this implies that $\eta = \d \vth$ is adapted to
$\pi:M \to B$.

The simplest occurrence of the mechanism described in abstract
terms in section 6.2 is for $M = \R^5$ with euclidean metric
and $B = \R^2$, i.e. $n=5$ and $k=2$\footnote{We need $1 < k <
n-2$: in the case $k=1$ the variational principle identifies a
vector field in a standard way, and $k = n-2$ gives a maximal
degree variational principle.}. We will analyze this case in full
detail. We will take coordinates $(x^1,x^2;z^1,z^2,z^3)$ in
$\R^5$; the space $B$ will correspond to the $(x^1,x^2)$ (i.e.
$B\ss M$ is given by $z^1=z^2=z^3=0$), so that the $z$ represent
coordinates along the fiber, i.e. vertical ones.

We write $d \vth = \eta \in \Lambda^3 (M)$ in the form $ \eta \ =
\a_1 \w \a_2 \w \a_3 $. We choose (see the discussion in sect.6)
$$ \a_a \ = \ \d z^a \, + \, B_{aj} \, \d x^j \ . $$
Thus the explicit expression for $\eta$ is:
$$ \begin{array}{rl}
\eta \ =& \d z^1 \w \d z^2 \w \d z^3 + \\
& + B_{31}\d z^1 \w \d z^2 \w \d x^1 + B_{32}\d z^1 \w \d z^2 \w \d x^2 + \\
& + B_{21}\d z^3 \w \d z^1 \w \d x^1 + B_{22}\d z^3 \w \d z^1 \w \d x^2 + \\
& + B_{11}\d z^2 \w \d z^3 \w \d x^1 + B_{12}\d z^2 \w \d z^3 \w \d x^2 + \\
& + (B_{21}B_{32}-B_{22}B_{31}) \d x^1 \w \d x^2 \w \d z^1 + \\
& + (B_{12}B_{31}-B_{11}B_{32}) \d x^1 \w \d x^2 \w \d z^2 + \\
& + (B_{11}B_{22}-B_{12}B_{21}) \d x^1 \w \d x^2 \w \d z^3 \ .
\end{array} $$

In this case, the variational ideal $\J (\vth,\pi) $ is generated
by the three $2$-forms $\psi_a:= (\pa / \pa z^a) \interno \eta $.
They are given explicitly by
$$ \begin{array}{rl}
\psi_1 \ =& \d z^2 \w \d z^3+B_{31}\d z^2 \w \d x^1+B_{32} \d z^2 \w \d x^2
-B_{21} \d z^3 \w \d x^1 + \\
& - B_{22} \d z^3  \w \d x^2 +(B_{21}B_{32}-B_{22}B_{31}) \d x^1 \w \d x^2 \ , \\
\psi_2 \ =& \d z^3 \w \d z^1+B_{11}\d z^3 \w \d x^1+B_{12} \d z^3 \w \d x^2
-B_{31} \d z^1 \w \d x^1 + \\
& - B_{32} \d z^1  \w \d x^2 + (B_{12}B_{31} - B_{11}B_{32}) \d x^1 \w \d x^2 \ , \\
\psi_3 \ =& \d z^1 \w \d z^2+B_{21}\d z^1 \w \d x^1+B_{22} \d z^1 \w \d x^2
-B_{11} \d z^2 \w \d x^1 + \\
& - B_{12} \d z^2  \w \d x^2 + (B_{11}B_{22}-B_{12}B_{21}) \d x^1 \w \d x^2 \ .
\end{array} $$

Let us now consider a section $\phi$, described in coordinates by
$z^a = z^a (x^1,x^2)$. We write $\phi^* (\psi_a) := \De_a \d x^1 \w
 \d x^2 \ = \ \De_a \om$, and $\phi$ is critical if and only if the $z^a (x^1,x^2)$ satisfy the equations $\De_a = 0$ for $a=1,2,3$.

By explicit computations, and writing $\pa_i := \pa / \pa x^i$, we obtain that these equations are:
$$ \begin{array}{ll}
\De_1 :=& (\pa_1 z^2) (\pa_2 z^3) - (\pa_2 z^2) (\pa_1 z^3) - B_{31} (\pa_2 z^2) +B_{32} (\pa_1 z^2) + \\
& + B_{21} (\pa_2 z^3) -B_{22} (\pa_1 z^3) + B_{21}B_{32}-B_{22}B_{31} \ = \ 0 \ , \\
\De_2 :=&  (\pa_2 z^1) (\pa_1 z^3) - (\pa_1 z^1) (\pa_2 z^3) +B_{31} (\pa_2 z^1) -B_{32} (\pa_1 z^1) + \\
& +B_{12} (\pa_1 z^3) -B_{11} (\pa_2 z^3)
+B_{12}B_{31}-B_{11}B_{32} \ = \ 0\ , \\
\De_3 :=&  (\pa_1 z^1) (\pa_2 z^2) - (\pa_2 z^1) (\pa_1 z^2) -B_{21} (\pa_2 z^1) +B_{22} (\pa_1 z^1) + \\
& +B_{11} (\pa_2 z^2) -B_{12}(\pa_1 z^2)
+B_{11}B_{22}-B_{12}B_{21} \ = \ 0 \ .
\end{array} $$

These equations can also be obtained (see theorem 2 and remark 7)
by requiring that ${\rm rank} (P) < 2$, with $P$ the matrix given
by
$$
P \ =\ \pmatrix{
B_{11}+ (\pa_1 z^1) & B_{12}+ (\pa_2 z^1) \cr
B_{21}+ (\pa_1 z^2) & B_{22}+ (\pa_2 z^2) \cr
B_{31}+ (\pa_1 z^3) & B_{32}+ (\pa_2 z^3) \cr}
$$
Indeed, $\De_a$ is the determinant of the matrix $\^P_a$ obtained from  $P$ by elimination of its $a$-th row.

Note also that introducing the vector fields
$$ X_i = {\pa \over \pa x^i} + B_{ai} {\pa \over \pa z^a} $$
the matrix $P$ is rewritten as
$$ P \ = \ \pmatrix{ X_1 (z^1) & X_2 (z^1 ) \cr
X_1 (z^2) & X_2 (z^2 ) \cr
X_1 (z^3) & X_2 (z^3 ) \cr} \ , $$
and the condition ${\rm rank} (P) < 2$ reads
$$ \epsilon_{ijk} X_1 (z^j) X_2 (z^k) \ = \ 0 \ \ \ i = 1,2,3 \ . $$

The characteristic distribution $ \D = D[\J (\vth,\pi)] $
associated to the variational ideal $\J (\vth,\pi) $ is given by
the vector field $Y$ on $M$ such that $Y \interno \psi_a = 0$ for
$a=1,2,3$; by lemma 1, $\D$ coincides with $\N (\d \vth)$. We have
by explicit computation that $\N(\eta) $ is a 2-dimensional
integrable distribution generated by
$$ \begin{array}{rl}
Y_1 \ =& (\pa / \pa x^1 ) \, - \, \[ B_{11} (\pa / \pa z^1)
+ B_{21} (\pa / \pa z^2) + B_{31} (\pa / \pa z^3) \] \ , \\
Y_2 \ =& (\pa / \pa x^2) \, - \, \[ B_{12} (\pa / \pa z^1)
+ B_{22} (\pa / \pa z^2) + B_{32} (\pa / \pa z^3) \] \ .
\end{array} $$

Then, critical sections for the variational principle associated to
$\vth$ can be obtained as sections of $\pi:M \to B$ that are
integral manifold (in the sense of definition 3) for the
characteristic distribution $ \D = D[\J (\vth,\pi)] $ generated by
$Y_1$ and $Y_2$.

In  particular, let us consider a section $\phi$ of  $\pi: M \to
B$ given by $z^a = \phi^a (x^1,x^2)$, for $a=1,2,3$. It is
immediate to check that a vector field
$$ X \ = \ f^1 {\pa \over \pa z^1} \, + \, f^2 {\pa \over \pa z^2} \, + \, F^1 {\pa \over \pa z^1} \, + \, F^2 {\pa \over \pa z^2} \, + \,
F^3 {\pa \over \pa z^3} \eqno(26) $$
is tangent to the section
$\phi$ if and only if
$$ F^a \ = \ f^1 {\pa \phi^a \over \pa x^1} \, + \,
f^2 {\pa  \phi^a \over \pa x^2} \ \ \ \ \ a=1,2,3 \ . $$

Looking for sections of $\pi$ which are integral manifolds of the
characteristic distribution $\D$ is equivalent to requiring that
the vector field $X$ defined in (26) belongs to $\D$. By (26),
$X=f^1 Y_1 + f^2 Y_2$ if and only if
$$ \begin{array}{l}
(\pa \phi^1 / \pa x^1) f^1+ (\pa \phi^1 / \pa x^2) f^2=
-f^1B_{11}-f^2B_{12}, \\
(\pa \phi^2 / \pa x^1) f^1+ (\pa \phi^2 / \pa x^2) f^2=
-f^1B_{21}-f^2B_{22}\\
(\pa \phi^3 / \pa x^1) f^1+ (\pa \phi^3 / \pa x^2) f^2=
-f^1B_{31}-f^2B_{32}
\end{array} $$

It is a trivial computation to prove that these equations are
equivalent to $\Delta_a=0$, just eliminating the variables $f^i$.

Then, critical section for the variational principle defined by
$\vth$ can be obtained as integral manifold of the characteristic
distribution $ \D = D [\J (\vth,\pi)] $ generated by $Y_1$ and
$Y_2$.

\section{Example 2: non maximally characteristic case}

Let us now consider a non maximally characteristic case, i.e. a
case with $h \not=0 $ (see sect.6.3). The simplest such case is
obtained for $n=6$ and $k=3$, with $h = 2k+1-n = 1$.

Thus we consider as $M$ the euclidean $\R^6$ space, fibered over
$B = \R^3$. We denote by $(x^1,x^2,x^3)$ coordinates on $B$, and
by $(z^1,z^2,z^3)$ coordinates in the fibers $\pi^{-1} (b)$.
Proceeding according to our general discussion, we write
$$ \cases{ \a_a = \d z^a + B_{ak} \d x^k & ($a=1,2,3$), \cr
\a_4 = C_k \d x^k \ , & \cr} $$ and $\eta = \a_1 \w \a_2 \w \a_3
\w \a_4$. Note that $(\pa / \pa z^a) \interno \a_m = \de_{am} $.
Hence
$$ \begin{array}{rll}
\Psi_1 \ =& \a_2 \w \a_3 \w \a_4 \ =& \chi_1 \ , \\
\Psi_2 \ =& - \a_1 \w \a_3 \w \a_4 \ =& \chi_2 \  , \\
\Psi_3 \ =&  \a_1 \w \a_2 \w \a_4 \ =& \chi_3 \  . \end{array} $$

In considering the pullbacks $\phi^* (\Psi_a)$, it is convenient
to introduce $\om = \d x^1 \w \d x^2 \w \d x^3$ and write $ \phi^*
(\Psi_a) = \De_a \cdot \om$. Note that
$$ \cases{\phi^*(\a_a ) = (\pa z^a / \pa x^k + B_{ak} ) \d x^k \ := \
F_{ak} \d x^k & for $a=1,2,3$, \cr \phi^* (\a_4) = C_k \d x^k \ .
& \cr}
$$ Therefore, with standard algebra,
$$ \phi^* (\Psi_a) \ = \ (1/2) \ \epsilon_{abc} \ F_{b \mu}
\, F_{c \nu} \, C_\s \ \d x^\mu \w \d x^\nu \w \d x^\s $$ and
hence, omitting a constant (1/12) factor,
$$ \De_a \ = \ \epsilon_{abc} \, \epsilon_{\mu \nu \s} \, F_{b \mu} \, F_{c \nu} C_\s \ . $$

The equations $\De_a = 0$ for $a=1,2,3$ can also be written in terms of the matrix
$$ P \ = \ \pmatrix{
B_{11} + \pa_1 z^1 & B_{12} + \pa_2 z^1 & B_{13} + \pa_3 z^1 \cr
B_{21} + \pa_1 z^2 & B_{22} + \pa_2 z^2 & B_{23} + \pa_3 z^2 \cr
B_{31} + \pa_1 z^3 & B_{32} + \pa_2 z^3 & B_{33} + \pa_3 z^3 \cr
C_1 & C_2 & C_3 \cr} $$ as the requirement that all the
three-dimensional submatrices $\^P_a$ obtained deleting from $P$
the $a$-th row, for $a=1,2,3$ have zero determinant. Note this
does not set any requirement on $\^P_4$.

We can rewrite the matrix $P$ in terms of three vector fields,
transversal to the fibers of $\pi : M \to B$ and defined as
$$ X_i \ := \ {\pa \over \pa x^i} \ + \ B_{ai} \, {\pa \over \pa z^a}
\ \ \ \ (i=1,2,3) \ . $$
With these, $P$ is rewritten as
$$ P \ = \ \pmatrix{
X_1 (z^1) & X_1 (z^2) & X_1 (z^3) \cr X_2 (z^1) & X_2 (z^2) & X_2
(z^3) \cr X_3 (z^1) & X_3 (z^2) & X_3 (z^3) \cr C_1 & C_2 & C_3
\cr} \ . $$

The equations $\De_a = 0$ are then written as
$$ \begin{array}{rl}
\De_1 \ :=& \
\[ X_2 (z^2) X_3 (z^3) - X_2 (z^3) X_3 (z^2) \] C_1 \\ & +
\[ X_2 (z^3) X_3 (z^1) - X_2 (z^1) X_3 (z^3) \] C_2 \\ & +
\[ X_2 (z^1) X_3 (z^2) - X_2 (z^2) X_3 (z^1) \] C_3 \ ; \\
\De_2 \ :=& \
\[ X_1 (z^2) X_3 (z^3) - X_1 (z^3) X_3 (z^2) \] C_1 \\ & +
\[ X_1 (z^3) X_3 (z^1) - X_1 (z^1) X_3 (z^3) \] C_2 \\ & +
\[ X_1 (z^1) X_3 (z^2) - X_1 (z^2) X_3 (z^1) \] C_3 \ ; \\
\De_3 \ :=& \
\[ X_1 (z^2) X_2 (z^3) - X_1 (z^3) X_2 (z^2) \] C_1 \\ & +
\[ X_1 (z^3) X_2 (z^1) - X_1 (z^1) X_2 (z^3) \] C_2 \\ & +
\[ X_1 (z^1) X_2 (z^2) - X_1 (z^2) X_2 (z^1) \] C_3 \ . \end{array} $$

Let us now consider $\N (\eta)$. Writing generic vector fields in
the form
$$ Y \ = \ f^i {\pa \over \pa x^i } \ + \ F^a {\pa \over \pa z^a} \ , $$
these are in $\N (\eta )$ if the coefficients satisfy the relations
$$ F^a \ = \ - B_{ai} f^i \ \ , \ \ C_k f^k \ = \ 0 \ . $$
The vector fields satisfying these conditions form a two
dimensional module; we can take as generators of $\N (\eta )$ e.g.
the vector fields
$$ \begin{array}{rl}
Y_1 \ =& \ C_3 (\pa / \pa x^1 ) \, - \, C_1 (\pa / \pa x^3) \ + \
[ B_{a3} C_1 - B_{a1} C_3 ] (\pa / \pa z^a ) \ , \\
Y_2 \ =& \ C_3 (\pa / \pa x^2 ) \, - \, C_2 (\pa / \pa x^3) \ + \
[ B_{a3} C_2 - B_{a2} C_3 ] (\pa / \pa z^a ) \ . \end{array} $$

\section{Example 3: non proper variational principle}

As a third and final example we will consider a case where $\d \vth$ admits an  annihilating vertical vector field, i.e. where the variational principle defined by $\vth$ is non proper, and $\eta$ not adapted to the fibration $\pi : M \to B$. We will of course require that $\eta$ is compatible with $\pi : M \to B$.

The simplest such case of interest in the present context is obtained for $k=2$ and $n=6$; note here $n > 2k +1$.

We will take coordinates $(x^1,x^2;z^1,z^2,z^3, w)$ in euclidean $\R^6$; the $(x^1,x^2)$ will be coordinates in the space $B$, and the
$(z^a,w)$ represent coordinates along the fibers, i.e. vertical ones.

Consider a form $\vth \in \La^2 (M)$ such that $\eta = \d \vth$ is nondegenerate and decomposable; we write it as
$ \eta \ = \a_1 \w \a_2 \w \a_3 $, and choose (see sect.7)
$$ \begin{array}{rl}
\a_1 \ = \ \d z^1 + B_{1k} \d x^1+B_{12} \d x^2 +C_1 \d w \ , \\
\a_2 \ = \ \d z^2+B_{21} \d x^1+B_{22} \d x^2 +C_2 \d w \ , \\
\a_3 \ = \ \d z^3+B_{31} \d x^1+B_{32} \d x^2 +C_3 \d w \ . \end{array}
$$

Then we have the following explicit expression for $\eta$:
$$ \begin{array}{rl}
\eta \ =& \d z^1 \w \d z^2 \w \d z^3 + B_{31}\d z^1 \w \d z^2 \w \d x^1 + B_{32}\d z^1 \w \d z^2 \w \d x^2 \\
& + C_3 \d z^1 \w \d z^2 \w \d w + B_{21}\d z^3 \w \d z^1 \w \d x^1
+ B_{22}\d z^3 \w \d z^1 \w \d x^2 \\
& + C_2 \d z^3 \w \d z^1 \w \d w + B_{11}\d z^2 \w \d z^3 \w \d x^1
+ B_{12}\d z^2 \w \d z^3 \w \d x^2 \\
& + C_1 \d z^2 \w \d z^3 \w \d w + (B_{21}B_{32}-B_{22}B_{31}) \d x^1 \w \d x^2 \w \d z^1 \\
& + (B_{12}B_{31}-B_{11}B_{32}) \d x^1 \w \d x^2 \w \d z^2
+ (B_{11}B_{22}-B_{12}B_{21}) \d x^1 \w \d x^2 \w \d z^3 \\
& + (C_1B_{21}-C_2 B_{11})\d x^1 \w \d z^3\ \d w +
(C_1 B_{22}-C_2 B_{12}) \d x^2 \w \d z^3 \w \d w \\
& + (C_2 B_{32}-C_3 B_{22}) \d x^2 \w \d z^1 \w \d w +
(C_3 B_{12}-C_1 B_{32}) \d x^2 \w \d z^2 \w \d w \\
& + (C_2 B_{31}-C_3 B_{21}) \d x^1 \w \d z^1 \w \d w +
(C_3 B_{11}-C_1 B_{31}) \d x^1 \w \d z^2 \w \d w \\
& + ( C_1 B_{32}B_{21} + C_3 B_{11}B_{22}+ C_2 B_{31}B_{12}- C_1
B_{22}B_{31} \\
& -C_2 B_{32}B_{11} -C_3 B_{12}B_{31}) \d x^1 \w \d x^2 \w \d w.
\end{array}
$$

In this case, the variational ideal $\J (\vth,\pi) $ is generated
by the three $2$-forms $\psi_a:= (\pa / \pa z^a) \interno \eta $,
as $\psi_4$ will be a linear combination of these, and more
precisely $\psi_4 = C_a \psi^a$.

We have indeed
$$ \begin{array}{rl}
\psi_1 =& \d z^2 \w \d z^3+B_{31}\d z^2 \w \d x^1+B_{32} \d z^2 \w \d x^2 +C_3 \d z^2\w \d w \\
& -B_{21} \d z^3 \w \d x^1 - B_{22} \d z^3  \w \d x^2
-C_2 \d z^3 \w \d w \\
& + (B_{21}B_{32}-B_{22}B_{31}) \d x^1 \w \d x^2 \\
& -(C_2 B_{32}-C_3 B_{22}) \d x^2 \w \d w
-(C_2 B_{31}-C_3 B_{21}) \d x^1 \w \d w \\
\psi_2 =& \d z^3 \w \d z^1+B_{11}\d z^3 \w \d x^1+B_{12} \d z^3 \w \d x^2
+C^1 \d z^3\w \d w \\
& -B_{31} \d z^1 \w \d x^1 - B_{32} \d z^1  \w \d x^2
-C^3 \d z^1\w \d w \\
& + (B_{11}B_{32}-B_{12}B_{31}) \d x^1 \w \d x^2 \\
& -(C_3 B_{12}-C_1 B_{32}) \d x^2 \w \d w
-(C_3 B_{11}-C_1 B_{31}) \d x^1 \w \d w \\
\psi_3 =& \d z^1 \w \d z^2+B_{21}\d z^1 \w \d x^1+B_{22} \d z^1 \w \d x^2 +C_2 \d z^1 \w \d w \\
& -B_{11} \d z^2 \w \d x^1 - B_{12} \d z^2  \w \d x^2 - C_1 \d z^2 \w \d w \\
& + (B_{11}B_{22}-B_{12}B_{21}) \d x^1 \w \d x^2 \\
& -( C_1 B_{21}-C_2 B_{11}) \d x^1 \w \d w -( C_1 B_{22}-C_2
B_{12}) \d x^2 \w \d w \ , \end{array} $$ while $\psi_4$ is given
by
$$ \begin{array}{rl}
\psi_4 =& C_3 \d z^1 \w \d z^2 + C_2 \d z^3 \w \d z^1 + C_1 \d z^2
\w \d z^3 \\
& + ( C_1 B_{21} - C_2 B_{11} ) \d x^1 \w \d z^3 + ( C_1 B_{22} -
C_2 B_{12} ) \d x^1 \w \d z^3 \\
& + ( C_2 B_{32} - C_3 B_{22} ) \d x^2 \w \d z^1 + ( C_2 B_{31} -
C_3 B_{21} ) \d x^1 \w \d z^1 \\
& + ( C_3 B_{12} - C_1 B_{32} ) \d x^2 \w \d z^2 + ( C_3 B_{11} -
C_1 B_{31} ) \d x^1 \w \d z^2 \\
& + [ C_1 (B_{21} B_{32} - B_{31} B_{22} ) + C_2 (B_{31} B_{12} \\
& - B_{11} B_{32} ) + C_3 ( B_{11} B_{22} - B_{12} B_{21} ) ] \ .
\end{array} $$

The equations $\phi^* (\psi_a) := \De_a \om = 0$ can be written in
terms of the matrix
$$ P \ = \ \pmatrix{
B_{11} + \pa_1 z^1 + C_1 \pa_1 w & B_{12} + \pa_2 z^1 + C_1 \pa_2 w \cr
B_{21} + \pa_1 z^2 + C_2 \pa_1 w & B_{22} + \pa_2 z^2 + C_2 \pa_2 w \cr
B_{31} + \pa_1 z^3 + C_3 \pa_1 w & B_{32} + \pa_2 z^3 + C_3 \pa_2 w \cr} $$
as the requirement that all the
two-dimensional submatrices $\^P_a$ obtained deleting from $P$
the $a$-th row, for $a=1,2,3$ have zero determinant.

Let us now consider $\N (\d \vth ) $. This is generated by the vector fields
$$ \begin{array}{rl}
Y_1 \ =& (\pa / \pa x^1 ) \, - \, \[ B_{11} (\pa / \pa z^1)
+ B_{21} (\pa / \pa z^2) + B_{31} (\pa / \pa z^3) \] \ , \\
Y_2 \ =& (\pa / \pa x^2) \, - \, \[ B_{12} (\pa / \pa z^1)
+ B_{22} (\pa / \pa z^2) + B_{32} (\pa / \pa z^3) \] \ , \\
Y_3 \ =& (\pa / \pa w) \, - \, \[ C_1 (\pa / \pa z^1) + C_2 (\pa /
\pa z^2) + C_3 (\pa / \pa z^3) \] \ ; \end{array} $$ note that
$Y_3$ is vertical for $\pi$.

The integral manifolds of $\D = \N (\d \vth)$ will be
three-dimensional. We are actually interested in integral
manifolds for the variational ideal $\J$ which are sections for
the bundle $(M,\pi,B)$ (we call these critical sections for
short); this means in particular that they are two dimensional and
transversal to fibers of $\pi$. Note this means that they are not
maximal integral manifolds for $\D$, at difference with the cases
considered before.

When we try to determine critical sections making use of our
knowledge of $\N (\d \vth )$, we should consider general sections
$\phi \in \Ga (\pi)$, i.e. manifolds $\{x,z,w\}$ identified by
$z^a = \phi^a (x^1,x^2)$ and $w = \phi^4 (x^1 , x^2)$, and require
that vector fields $X$ which are tangent to $\phi$ are in the
distribution $\D = \N (\d \vth )$. If this is the case, the section $\phi$ is indeed a critical section.

Proceeding in this way, we write a general $X$ in the form
$$ X \ = \ f^i {\pa \over \pa x^i } \ + \ F^a {\pa \over \pa
z^a} \ + \ F^4 {\pa \over \pa w } \ ; \eqno(27) $$ this is tangent
to the section $\phi$ if and only if
$$ F^a \ = \ {\pa \phi^a \over \pa x^i } \ f^i \ . \eqno(28) $$
A vector field $X$ in the form (27) and satisfying (28) is in $\D$
if
$$ \begin{array}{rl}
(\pa \phi^1 / \pa x^1) f^1 + (\pa \phi^1 / \pa x^2) f^2 \ = &
-  B_{11} f^1 - B_{12} f^2 - C_1 f^1 (\pa \phi^4 / \pa x^1) \\
 & - C_1 f^2 (\pa \phi^4 / \pa x^2) , \\
(\pa \phi^2 / \pa x^1) f^1 + (\pa \phi^2 / \pa x^2) f^2 \ = &
- B_{21} f^1 - B_{22} f^2 - C_2 f^1 (\pa \phi^4 / \pa x^1) \\
 & - C_2 f^2 (\pa \phi^4 / \pa x^2) \\
(\pa \phi^3 / \pa x^1) f^1 + (\pa \phi^3 / \pa x^2) f^2 \ = & -
B_{31} f^1 - B_{32} f^2 - C_3 f^1 (\pa \phi^4 / \pa x^1) \\
 & - C_3
f^2 (\pa \phi^4 / \pa x^2) \ .
\end{array} $$

It is a simple matter to check that, eliminating the variables $f^i$ from this system, we recover the equations $\De_a = 0$ ($a=1,2,3$).

\vfill\eject

\end{document}